\documentclass[
reprint,
showkeys,
aps,
prd,
longbibliography,
nofootinbib,
twocolumn
]{revtex4-2}
\usepackage[utf8]{inputenc}
\usepackage{amsmath}
\usepackage{amssymb}
\usepackage{xstring}
\usepackage{siunitx}
\usepackage{slashed}
\usepackage{dsfont}
\usepackage{upgreek}
\usepackage{bm}
\usepackage{mathtools}

\usepackage[hyperref,dvipsnames]{xcolor}
\definecolor{goethe-blau}{cmyk}{1.0,0.2,0.0,0.4}
\definecolor{hellgrau}{cmyk}{0.04,0.04,0.05,0.02}
\definecolor{sandgrau}{cmyk}{0.12,0.09,0.13,0.0}
\definecolor{dunkelgrau}{cmyk}{0.25,0.25,0.30,0.75}
\definecolor{emo-rot}{cmyk}{0.04,1.0,0.8,0.07}
\definecolor{purple}{cmyk}{0.08,1.0,0.3,0.36}
\definecolor{senfgelb}{cmyk}{0.01,0.25,1.0,0.05}
\definecolor{gruen}{cmyk}{0.62,0.4,0.87,0.09}
\definecolor{magenta}{cmyk}{0.08,0.86,0.12,0.12}
\definecolor{orange}{cmyk}{0.0,0.7,1.0,0.04}
\definecolor{sonnengelb}{cmyk}{0.0,0.12,0.95,0.0}
\definecolor{helles-gruen}{cmyk}{0.4,0.17,0.81,0.07}
\definecolor{lichtblau}{cmyk}{0.8,0.0,0.06,0.04}

\usepackage[
colorlinks,
pdfpagelabels,
breaklinks,
pdfstartview=FitH,
bookmarksopen=true,
bookmarksnumbered=true,
bookmarksopenlevel=2,
plainpages=false,
hypertexnames=false,
citecolor=emo-rot,
linkcolor=goethe-blau,
urlcolor=purple,
pdftitle={Regularization effects in the Nambu-Jona-Lasinio model: Strong scheme dependence of inhomogeneous phases and persistence of the moat regime}
,pdfauthor={Laurin Pannullo, Marc Wagner, Marc Winstel},
pdfkeywords={Nambu-Jona-Lasinio model, stability analysis, inhomogeneous phases, moat regimes, regulator dependence, mean-field}
]{hyperref}

\usepackage{bookmark}
\usepackage[capitalise]{cleveref}
\usepackage{multirow}
\usepackage{xstring}
\usepackage{graphicx}	
\usepackage{dcolumn}	

\usepackage{xparse}
\ExplSyntaxOn
\DeclareExpandableDocumentCommand{\IfNoValueOrEmptyTF}{mmm}
{
	\IfNoValueTF{#1}{#2}
	{
		\tl_if_empty:nTF {#1} {#2} {#3}
	}
}
\ExplSyntaxOff

\NewDocumentCommand{\symbolWithOptionalSubScript}{o m} {%
	\IfNoValueOrEmptyTF {#1} {%
		#2
	}{%
		#2_{#1}
	}%
}

\NewDocumentCommand{\symbolWithOptionalArgument}{o m} {%
	\IfNoValueOrEmptyTF {#1} {%
		#2
	}{%
		#2\left(#1\right)
	}%
}

\DeclareMathOperator{\arctantwo}{arctan2}

\DeclareMathOperator{\arsinh}{arsinh}
\DeclareMathOperator{\artanh}{artanh}

\DeclareMathOperator{\arcoth}{arcoth} 

\DeclareMathOperator*{\argmin}{arg\,min}
\newcommand{\argPlaceholder}{\cdot}
\DeclareMathOperator{\Tr}{\mathrm{Tr}}
\DeclareMathOperator{\tr}{\mathrm{tr}}
\newcommand{\vecb}{\mathbf}
\newcommand{\X}[1]{#1}

\newcommand{\x}[1]{\vecb{#1}}
\newcommand{\abs}[1]{\lvert#1\rvert}

\newcommand{\Abs}[1]{\left\lvert#1\right\rvert}
\DeclarePairedDelimiter\floor{\lfloor}{\rfloor}
\newcommand{\ifc}{\text{if }}
\newcommand{\otherwisec}{\text{otherwise }}
\newcommand{\iu}{\mathrm{i}}
\providecommand{\eu}{\mathrm{e}}

\newcommand{\N}{N}
\newcommand{\Nf}{N_f}
\newcommand{\Nc}{N_c}
\newcommand{\Ng}{N_{\gamma}}

\newcommand{\homPot}{\ensuremath{U}}

\NewDocumentCommand{\twopid}{ o } {%
	\IfNoValueTF {#1} {%
		2\uppi%
	}
	{%
		(2\uppi)^{#1}%
	}%
}

\NewDocumentCommand{\intMeasureOverPiNew}{ o m } {%
	\tfrac{\dr[#2][#1]}{\twopid[#1]}\,
}
\newcommand{\kron}[2]{\ensuremath{\delta_{#1,#2}}}

\newcommand{\E}[1]{E_{#1}}

\newcounter{numrefs}
\makeatletter
\newcommand{\Rcite}[1]{%
	\setcounter{numrefs}{0}
	\@for\@temp:=#1\do{\stepcounter{numrefs}}
	\ifnum\value{numrefs}>1%
	Refs.~\cite{#1}%
	\else%
	Ref.~\cite{#1}%
	\fi%
}
\makeatother

\NewDocumentCommand{\distributionF}{ o m } {
	\IfNoValueOrEmptyTF {#1} {%
		n\left(#2\right)%
	}
	{%
		n^{#1}\!\left(#2\right)%
	}
}

\NewDocumentCommand{\distributionFBar}{ o m } {
	\IfNoValueOrEmptyTF {#1} {%
		\bar{n}\left(#2\right)%
	}
	{%
		\bar{n}^{#1}\!\left(#2\right)%
	}
}

\newcommand{\OneMinusnsDef}[1]{1-\distributionF{#1}-\distributionFBar{#1}}

\newcommand{\mstar}{\bar{\sigma} }
\newcommand{\mstarMin}{\bar{\Sigma} }

\newcommand{\Ell}{L}

\newcommand{\ellZero}[1][ ]{l_{0#1}}
\newcommand{\ellOne}[1][ ]{l_{1#1}}
\newcommand{\ellTwo}[1][ ]{l_{2#1}}

\newcommand{\ellOneArgs}[3]{\ellOne{}\!\left(#1,#2,#3\right)}

\newcommand{\EllTwo}[1][]{\Ell_{2#1}}

\newcommand{\gtwo}[1][]{\ensuremath{\Gamma^{(2)}_{#1}}}
\newcommand{\gtwoArgs}[5][]{\ensuremath{\gtwo[#1]\left(#2,#3,#4,#5\right)}}

\newcommand{\z}[1][]{\ensuremath{z_{#1}}}
\newcommand{\zArgs}[4][]{\z[#1]\left(#2,#3,#4\right)}

\newcommand{\zPhys}[1][]{\ensuremath{Z_{#1}}}

\newcommand{\qChar}{\ensuremath{Q}}

\makeatletter
\NewDocumentCommand{\dr}{ o o } {%
	\mathop{}\!\mathrm{d}%
	\IfNoValueTF {#2} {%
	}
	{%
		^{#2}\!%
	}
	\IfNoValueTF {#1} {%
	}
	{%
		#1\,%
	}%
}

\NewDocumentCommand{\Dr}{ o o } {%
	\mathop{}\!\mathcal{D}%
	\IfNoValueTF {#2} {%
	}
	{%
		^{#2}\!%
	}
	\IfNoValueTF {#1} {%
	}
	{%
		#1\,%
	}%
}
\makeatother
\newcommand{\BetafunctionWOArgs}{%
	B%
}

\NewDocumentCommand{\Betafunction}{ o m m} {%
	\IfNoValueTF {#1} {%
		\BetafunctionWOArgs\left(#2, #3\right)%
	}{%
		\BetafunctionWOArgs\left(#1; #2, #3\right)%
	}%
}

\DeclareMathOperator{\Det}{\mathrm{Det}}

\providecommand{\iu}{\mathrm{i}} 
\newcommand{\e}{\mathrm{e}}   
\newcommand{\seff}{S_{\text{eff}}}

\newcommand{\Nbar}{\bar{N}}
\renewcommand{\Nf}{N_f}

\newcommand{\Mass}{M}
\newcommand{\Mzero}{\Mass_0}

\newcommand{\regulator}{\Lambda}
\newcommand{\regulatorSC}{\regulator_{\mathrm{\acrshort{sc}}}}
\newcommand{\regulatorPV}{\regulator_{\mathrm{\acrshort{pv}}}}
\newcommand{\regulatorLFT}{\regulator_{\mathrm{\acrshort{lft}}}}
\newcommand{\NPV}{N_{\text{PV}}}
\newcommand{\pvsum}[1][ ]{\sum_{k{=#1}}^{\NPV}}
\newcommand{\pvsumck}[1][0]{\pvsum[#1] c_k}

\newcommand{\Hom}[1]{\ensuremath{\bar{#1}}}
\newcommand{\HomMin}{\ensuremath{\Hom{\Sigma}}}
\newcommand{\quarkPropagatorSLAC}[1][]{\symbolWithOptionalArgument[#1]{\quarkPropagatorSymbol_{\mathrm{SLAC}}}}
\newcommand{\quarkPropagatorHybrid}[1][]{\symbolWithOptionalArgument[#1]{\quarkPropagatorSymbol_{\mathrm{Hybrid}}}}
\newcommand{\fermionDisp}{\mathcal{P}}
\newcommand{\fermionDispSLAC}{\fermionDisp_\mathrm{SLAC}}

\newcommand{\hybridTheta}{Hybrid$_\Theta$}
\newcommand{\hybridCos}{Hybrid$_{\text{cos}}$}

\newcommand{\weightFunctionSymbol}{W}

\newcommand{\weightFunctionFourier}[1][]{\tilde{\weightFunctionSymbol}_{#1}}

\newcommand{\weightFunctionSymbolOneD}{w}

\newcommand{\weightFunctionFourierOneD}[1][]{\tilde{\weightFunctionSymbolOneD}_{#1}}
\newcommand{\weightFunctionFourierCosOneD}{\weightFunctionFourierOneD[\cos]}
\newcommand{\weightFunctionFourierHeavisideOneD}{\weightFunctionFourierOneD[\Theta]}

\newcommand{\Gcoupling}{G}
\newcommand{\vertexV}{c}
\newcommand{\piondecay}{f_\pi}
\newcommand{\mub}{\bar{\mu}}

\newcommand{\quarkPropagatorSymbol}{S}

\usepackage[acronym]{glossaries-extra}
\glssetcategoryattribute{acronym}{nohyperfirst}{true}
\setabbreviationstyle[acronym]{long-short}
\setabbreviationstyle[ignored]{long-noshort}

\newacronym{uv}{UV}{ultra-violet}
\newacronym{njl}{NJL}{Nambu-Jona-Lasinio}
\newacronym{gn}{GN}{Gross-Neveu}
\newacronym{qcd}{QCD}{Quantum Chromodynamics}
\newacronym{ff}{FF}{four-fermion}
\newacronym{hbp}{HBP}{homogeneously broken phase}
\newacronym{ip}{IP}{inhomogeneous phase}
\newacronym{sp}{SP}{symmetric phase}
\newacronym{cep}{CEP}{critical endpoint}
\newacronym{lp}{LP}{Lifshitz point}
\newacronym{sc}{SC}{spatial momentum cutoff}
\newacronym{pv}{PV}{Pauli-Villars}
\newacronym{lft}{LFT}{lattice field theory}
\newacronym{nmr}{NMR}{no medium regularization}
\newacronym{rs}{RS}{regularization scheme}

\usepackage{soul}

\begin{document}

	\title{
		Regularization effects in the Nambu-Jona-Lasinio model: Strong scheme dependence of inhomogeneous phases and persistence of the moat regime
	}
	
	\author{Laurin Pannullo$^{1}$}
	\email{lpannullo@physik.uni-bielefeld.de}
	\author{Marc Wagner$^{2,3}$}
	\email{mwagner@itp.uni-frankfurt.de}
	\author{Marc Winstel$^{2}$}
	\email{winstel@itp.uni-frankfurt.de}
	\affiliation{
		$^{1}$Fakultät für Physik, Universität Bielefeld,
		Universitätsstraße 25, \mbox{D-33615 Bielefeld, Germany}.
		\\
		$^{2}$Institut für Theoretische Physik, Goethe Universität Frankfurt am Main, Max-von-Laue-Straße 1, \mbox{D-60438 Frankfurt, Germany}
		\\
		$^{3}$Helmholtz Research Academy Hesse for FAIR,
		Campus Riedberg, Max-von-Laue-Straße 12
		, \mbox{D-60438 Frankfurt, Germany}
	}

	\date{\today}
	
	\begin{abstract}
	    This work investigates the phase structure of the non-renormalizable (3+1)-dimensional Nambu-Jona-Lasinio (NJL) model with particular focus on inhomogeneous phases (IPs), where the chiral condensate is non-uniform in space, and the closely related moat regimes, where mesonic dispersion relations favor non-vanishing momenta.
		We use the mean-field approximation and consider five different regularization schemes including three lattice discretizations.
		The results within the different regularization schemes are systematically analyzed in order to study the dependence of the IP on the choice of regulatization scheme and regulator value.
		The IP exhibits a drastic dependence on the chosen regularization scheme rendering any physical interpretation of results on inhomogeneous phases in this model doubtful.
		In contrast, we find only a mild scheme dependence of the moat regime suggesting	that its existence is a consequence of the action of the NJL model and its symmetries and, thus, that it might also exist in QCD.
		
	\end{abstract} 

	\maketitle
	\section{Introduction}
	
		\Glspl{ip} are phases where in addition to chiral symmetry also translational invariance is broken, i.e., the chiral condensate $\langle \bar{\psi} \psi \rangle$ is a function of the spatial coordinates $\vec{x} = \left(x_1, x_2, x_3\right)$. 
		An \gls{ip} is one of several, not necessarily mutually exclusive scenarios which are considered to be relevant in the phase diagram of \gls{qcd} at non-vanishing chemical potential $\mu$ and low to intermediate temperature $T$. 
		Other conjectured scenarios for the \gls{qcd} phase diagram in that region include color-superconducting phases \cite{Alford:2007xm}, quarkyonic matter \cite{McLerran:2007qj, Kojo:2009ha} or liquid-crystal like behavior \cite{ Lee:2015bva, Hidaka:2015xza}.
		Phases with spatially modulated chiral currents, that are observed in holographic models fitted to lattice \gls{qcd} data \cite{Nakamura:2009tf,Ooguri:2010kt,Ooguri:2010xs,Demircik:2024aig,CruzRojas:2024igr}, could possibly coincide or be an alternative scenario to an \gls{ip}.  
		
		While the phase diagram of \gls{qcd}, is well understood for small chemical potentials from lattice simulations \cite{Aoki:2006we,Borsanyi:2024wuq,Bellwied:2015rza,HotQCD:2018pds}, the phase structure for intermediate and large chemical potentials is less clear and an active area of current research.
		A common scenario is that the crossover is replaced at some non-vanishing value for $\mu$ by a first-order transition starting at a \gls{cep} and continuing down to zero temperature.
		Various functional methods predict this point at $\mu_{\mathrm{cep}} > T_{\mathrm{cep}}$ \cite{Fischer:2018sdj,Fu:2019hdw,Gao:2020fbl}. 
		While these methods necessitate certain approximations, their predictions are supported by lattice results that determine the location of the \gls{cep} by analyzing Lee-Yang zeros \cite{Clarke:2024seq}.
		Moreover, similar findings are obtained when using \gls{qcd}-inspired models such as the \gls{njl} model or the quark-meson model, see, e.g., \Rcite{Buballa:2003qv,Schaefer:2004en}.
		Due to the technical challenges for the lattice and functional methods at intermediate $\mu$, e.g., the sign problem, the majority of the current \gls{qcd} phase diagram predictions of a first order phase transition and a \gls{cep} are based on \gls{qcd}-inspired models.
			
		The above described scenario is based on the assumption of exclusively homogeneous condensation.
		When allowing the condensates to depend on the spatial coordinates, the first-order phase transition between the \gls{hbp} and the \gls{sp} is typically replaced and gets covered by an \gls{ip} in \gls{njl} and quark-meson model calculations \cite{Sadzikowski:2000ap,Nakano:2004cd, Nickel:2009wj,Carignano:2014jla, Buballa:2014tba}. 
		Often, the former \gls{cep} is then replaced by a so-called \gls{lp}, which coincides with the common end point of the two boundary lines of the \gls{ip} \cite{Nickel:2009wj,Buballa:2014tba, Carignano:2014jla}.
		Such model calculations have lead to the conjecture that the \gls{ip} might be a feature of the \gls{qcd} phase diagram at non-vanishing $\mu$ and also at large magnetic fields \cite{Ferrer:2019zfp, Gyory:2022hnv, Ferrer:2023xvl}.
		
		Apart from model investigations there is also evidence from functional approaches supporting this conjecture.
		Studies employing a Dyson-Schwinger approach to \gls{qcd} have shown that an \gls{ip} is a self-consistent solution in the \gls{qcd} phase diagram \cite{Muller:2013tya} and instabilities of a chirally symmetric solution towards an inhomogeneous condensate exist \cite{Motta:2023pks,Motta:2024agi}.
		Also, investigations using the Functional Renormalization Group \cite{Fu:2019hdw} support the existence of a moat regime in \gls{qcd}, a regime with an exotic meson dispersion relation featuring a minimum at non-vanishing momentum \cite{Pisarski:2021qof}.
		It is argued that this moat dispersion relation is a precursor to an \gls{ip}, since it favors particles with non-vanishing momenta.
		Indeed, the moat dispersion relation was recently shown to coincide with large parts of the \gls{ip} in the ($1+1)$-dimensional \gls{gn} model, but it can also exist without the presence of inhomogeneous condensates \cite{Koenigstein:2021llr}.
		
		The model calculations with focus on \glspl{ip} are predominantly restricted to the mean-field approximation, i.e., neglect bosonic quantum fluctuations. 
		There is an ongoing discussion whether \glspl{ip} persist when going beyond this approximation \cite{Lenz:2020bxk, Lenz:2020cuv, Stoll:2021ori,Lenz:2021kzo, Ciccone:2022zkg, Ciccone:2023pdk} or whether alternative scenarios such as liquid-crystal behavior \cite{Kolehmainen:1982jn,Hidaka:2015xza,Lee:2015bva, Akerlund:2016myr} or the so-called quantum pion liquid regime \cite{Schindler:2019ugo, Pisarski:2020dnx, Haensch:2023sig, Winstel:2024dqu, Winstel:2024qle} are realized through disordering by fluctuations of different types of Goldstone bosons.
		Typically, phases with symmetry breaking such as the \gls{hbp} and the \gls{ip} are weakened by quantum and/or thermal fluctuations \cite{Witten:1978qu,Hidaka:2015xza, Lenz:2020bxk, Stoll:2021ori, Melin:2024oee}. 
		However, these findings are either based on low-dimensional model calculations or involve drastic approximations. 
		Thus, clarifying the fate of the \gls{ip} beyond the mean-field approximation in theories with more relevance for phenomenology is an important future direction in the investigation of the \gls{qcd} phase diagram. 
		
		An obvious and necessary first step in this direction, is to study the \gls{rs} and regulator dependence of the \gls{ip} in mean-field theory before extending state of the art lattice field theory simulations of the \gls{ip} in \gls{njl}-type models \cite{Lenz:2020bxk, Lenz:2021kzo, Lenz:2023gsq} to $(3+1)$-dimensions.
		Recently, it has been demonstrated in the $(2+1)$-dimensional \gls{gn} model that the existence and shape of the \gls{ip} strongly depends on the used \gls{rs} and on the respective value of the regulator \cite{Buballa:2020nsi, Narayanan:2020uqt}.
		Moreover, in this model and also other related four-fermion models in $2+1$ dimensions, the \gls{ip} is only present at finite values of the regulator.
		The \gls{ip} is absent in the renormalized limit \cite{Pannullo:2023one}.
		This observation could be of particular relevance in the $(3+1)$-dimensional \gls{njl} model, which is non-renormalizable. 
		Its phase diagram obtained within different continuum \glspl{rs} was compared in \Rcite{Partyka:2008sv} with the main result that all three expected phases (\gls{sp}, \gls{hbp}, \gls{ip}) are present in the phase diagram, but with large quantitative differences.
		This study, however, compared the \glspl{rs} only at a single parameter set for each \gls{rs}, where central observables such as the constituent quark mass in the vacuum did not agree between the \glspl{rs}.
	
		In the present work, we significantly extend and improve this study in several ways.
		First, we determine the phase diagrams within different \glspl{rs} for a large range of constituent quark masses, which allows to compare in a more systematic and physically meaningful way.
		Second, we include three lattice discretizations as additional \glspl{rs} in the comparison.
		This is particularly interesting, since lattice Monte-Carlo simulations would be an ideally suited approach to study the phase diagram of this model beyond the mean-field approximation in a rigorous way.
		Moreover, we investigate the effect of excluding the medium part of the theory (which are finite without regularization) from the regularization. 
		This was already shown to have a strong effect on certain features of the \gls{njl} model and can, e.g., cure causality violations that otherwise arise at finite density within cutoff schemes \cite{Pasqualotto:2023hho}.
		Finally, we add a completely new direction and investigate the existence, extent and \gls{rs} dependence of the moat regime in the phase diagram at both zero and non-zero temperature and chemical potentials.
			
		This paper is organized as follows. 
		In \cref{sec:def_NJl} we recapitulate the definition of the \gls{njl} model and derive all relevant quantities for the study of its phase diagram. 
		Furthermore, we briefly discuss the five employed \glspl{rs} and the parameter matching, which is based on the quark constituent mass $\Mzero$ and the pion decay constant $\piondecay$.
		In \cref{sec:IPResults} we present the results regarding the \glspl{ip} including the phase diagram for each of the studied \glspl{rs} in the $\left(\Mzero, \mu \right)$-plane at zero temperature.
		Results for the moat regime for both $T = 0$ and $ T \neq 0$ are discussed in \cref{sec:moatres}.
		We conclude in \cref{sec:conclusions}.  
	
	\section{The Nambu-Jona-Lasinio model in $3+1$ dimensions}
	\label{sec:def_NJl}
	
		The action of the $\left(3+1\right)$-dimensional \gls{njl} model in Euclidean spacetime is 
		\begin{align}
		\label{eq:fermi_action}
		S[\bar{\psi},\psi] = \int_0^\beta \dr \tau \int \dr ^3 x \, \biggl\{& \bar\psi \left( \slashed \partial  
		+ \gamma_0 \mu \right) \psi +\\
		&{}+ G \left[\left( \bar\psi \psi\right)^2 + \left(\bar\psi \i \gamma_5 \pmb \tau \psi\right)^2 \right]\biggr\}, \nonumber 
		\end{align}
		where $\bar\psi$ and $\psi$ are fermion fields, describing massless quarks, with $\Nbar = \Nf \times \Nc \times \Ng = 2 \times 3 \times 4$ components (representing the number of quark flavors, colors and spin components, respectively).
		We use the slash notation $\slashed \partial = \gamma_\mu \partial_\mu$, where the matrices $\gamma_\mu$ fulfill the Clifford algebra for a Euclidean metric.
		The spacetime integration is restricted to $[0, \beta] \times \mathbb{R}^3$ with anti-periodic boundary conditions for the fermion fields in the imaginary time direction.
		The temporal extent $\beta$ is given by  the inverse temperature $1 / T$, $\mu$ is the quark chemical potential and $\pmb \tau$ is the vector of Pauli matrices acting in flavor space. 
		The coupling constant $G$ can be tuned such that certain vacuum observables agree with corresponding experimental values, e.g..~the pion decay constant (see \cref{sec:parameterfitting}). 
		
		The action \eqref{eq:fermi_action} is invariant under global $\mathrm{U}(1) \times \mathrm{SU}_V(2) \times \mathrm{SU}_A(2)$ transformations with the generators $\mathds{1}, \pmb{\tau}, \pmb{\tau} \gamma_5$, which is the chiral symmetry group in the massless limit of $2$-flavor \gls{qcd}.
		
		To get rid of the four-fermion term, one typically introduces bosonic auxiliary fields $\sigma$ and $\pmb \pi$ via a Hubbard-Stratonovich transformation.
		The resulting partition function with action
		\begin{align}
		\label{eq:partbos_action}
		S[\bar{\psi},\psi, \sigma, \pmb \pi] = \int \dr ^4 x \, \left\{ \bar\psi  D  \psi + \frac{\sigma^2 + \pmb{\pi}^2}{4G}\right\},
		\end{align}
		where
		\begin{equation}
		\label{eq:Dop} D = \slashed \partial+\gamma_0\mu  + \sigma + \iu \gamma_5\, \pmb \tau \cdot \pmb \pi,
		\end{equation}
		provides access to mesonic degrees of freedom in a more obvious way and is equivalent to the purely fermionic partition function.
		The expectation values of the bosonic fields satisfy the Ward identities
		\begin{align}
			\langle \bar{\psi} \psi \rangle = - \tfrac{1}{2G} \langle \sigma \rangle, \quad \langle \bar{\psi} \iu \gamma_5 \pmb{\tau} \psi \rangle = -\tfrac{1}{2G} \langle \pmb{\pi} \rangle,
		\end{align}
		and, thus, can be used as indicators for the spontaneous breaking of the chiral symmetry $\mathrm{SU}_A(2)$. The symmetry transformations $\mathrm{SU}_V(2) \times \mathrm{SU}_A(2)$ on the level of the purely fermionic action \eqref{eq:fermi_action} correspond to $\mathrm{O}(4)$ rotations of the field vector $\phi= (\sigma, \pmb \pi)$ for the action \eqref{eq:partbos_action}.
		
		As the fermion fields appear exclusively in a bilinear form in the action \eqref{eq:partbos_action}, the corresponding integration over Grassmann variables can be carried out, resulting in a purely bosonic path integral, 
		\begin{align}
		 Z &= \int \mathcal{D}\sigma \, \mathcal{D}\pmb \pi\, e^{-\seff[\sigma,\pmb \pi]} , \label{eq:part_func}
		\end{align}
		with the effective action 
		\begin{align}
			\label{eq:S_eff} \seff[\sigma, \pmb \pi] &=\int_0^\beta \dr \tau  \int \dr ^3 x \ \frac{\sigma^2 + \pmb \pi ^2}{4G}  - \Tr \ln\, D  . 
		\end{align}
		 
		We restrict this work to the mean-field approximation, where fluctuations of $\sigma$ and $\pmb{\pi}$ are suppressed.
		On the one hand this is a severe simplification compared to the full quantum field theory. 
		On the other hand the dependence on the fermionic \gls{uv}-regularization is still present within this approximation.
		The mean-field approximation is often considered as an appropriate starting point that is computationally less expensive than fully non-perturbative calculations with functional methods or lattice simulations.
		
		Within the mean-field approximation, the integration over $\sigma$ and $\pmb{\pi}$ in the partition function \eqref{eq:part_func} is reduced to the  contribution of a single field configuration, which is the global minimum of $\seff$ with respect to the bosonic fields, i.e., the  functional integration in the partition function is reduced to a minimization problem.
		In case of a spontaneously broken symmetry, one finds multiple degenerate minima.
		In the mean-field approximation, however, it is appropriate to simply pick one of the equivalent minima, see, e.g., \Rcite{Asakawa:1989bq, Buballa:2003qv, Buballa:2014tba, Heinz:2015lua, Islam:2023zpl} for corresponding discussions.\footnote{In case of a first order phase transition, one finds multiple degenerate minima, which are not connected by symmetry transformations.
		In this case we refrain from evaluating observables directly at the phase boundary line in order to avoid ambiguities.} 
		
		In this work, we restrict the bosonic fields to exhibit only a spatial, but not a temporal dependence, i.e., $\phi(x) = \phi(\vec{x})$, where $\vec{x} = (x_1, x_2, x_3)$.
		A further simplifying restriction, which we impose for a few specific calculations, is the restriction to homogeneous bosonic field $\sigma, \pmb{\pi} = \Hom{\sigma}, \Hom{\pmb{\pi}}$.
		In this particular case, the $\mathrm{O}(4)$ symmetry allows to set $\Hom{\pmb{\pi}} = 0$. 
		 
    \subsection{Homogeneous effective potential\label{sec:hompot}}
    
	    Evaluating the $\ln \Det D$ contribution in $\seff$ analytically is in general not possible for arbitrary field configurations.
	    While some specific $1$-dimensional modulations of the bosonic fields admit an expression of $\ln \Det D$ by knowledge of the density of states \cite{Basar:2008im,Basar:2008ki,Basar:2009fg,Nickel:2009wj}, one knows the exact eigenvalues for homogenoeus bosonic fields, as they coincide with eigenvalues of free fermions with the dynamically generated fermion mass $m = M = \sqrt{\Hom{\sigma}^2 + \Hom{\pmb{\pi}}^2}$.
	    The homogeneous, effective potential is defined as
	    \begin{equation}
	    	\homPot(M, \mu, T) = \frac{1}{\beta V} \seff[\Hom{\sigma}, \Hom{\pmb{\pi}}] \label{eq:hompot}
	    \end{equation}
		and can be rewritten into a compact form containing a simple momentum integral
		\begin{equation}
			\homPot(M, \mu, T) = \frac{\Mass^2}{4G}  - \frac{\Nbar}{2} \ellZero(M, \mu, T), \label{eq:hompotellZero}
		\end{equation}
		(see, e.g., \Rcite{Kapusta:2006pm, Wipf:2013vp}).
	    Here, $V$ is the spatial volume, for which we typically consider the limit $V \rightarrow \infty$ and $\ellZero$ is a \gls{uv}-divergent momentum integral that is defined in \cref{eq:ellZeroDef}.
	
	    If the ground state of the system, at given $\mu$ and $T$, is given by a homogeneous condensate, the homogeneous effective potential evaluated at its global minimum $M=\HomMin$ is equal to the thermodynamic grand potential, i.e., $\Omega(\mu, T) = \homPot(\HomMin, \mu, T)$. 
	    The configuration $\HomMin(\mu,T)$ can be determined using the gap equation
	    \begin{align}
	    &\frac{\partial \homPot}{\partial M}\bigg|_{M = \HomMin} = 0 \; \Rightarrow \, M \left( \frac{1}{2G} - \N \ell_1(M, \mu, T) \right)\bigg|_{M = \HomMin} = 0, \label{eq:gap}
	    \end{align}
	    where $\ellOne$ is a \gls{uv}-divergent integral, which requires regularization, and is defined in \cref{eq:ellOne}. 
	    If there is more than one solution of \cref{eq:gap}, one can determine the global minimum by inserting each of them in \cref{eq:hompotellZero} and comparing the corresponding values of $U$. \\
	    
	    The appearing divergences in $\homPot$ and $\ellOne$  are dependent on $M$ and cannot be compensated by $G$.
	    Thus, the regulator cannot be removed using a renormalization prescription as in renormalizable quantum field theories.
	    The results will instead depend on both the chosen \gls{rs} and the respective regulator $\Lambda$.
	    In other words, the \gls{rs} becomes to some extent part of the theory and $\Lambda$ can be understood as a parameter of the theory. 
	    We discuss more about the role of $\regulator$ in \cref{sec:parameterfitting}.   
	    More detailed formulas for $\homPot$ and $\ellOne$ can be found in \cref{app:formulas_hompout} and \cref{app:formulas_stability}, respectively.
    
    \subsection{Stability analysis}
    \label{sec:stability_analysis}
    
	    There are two common approaches to investigate inhomogeneous condensates.
	    Either one minimizes the effective action in the space of inhomogeneous field configurations to find the energetically favored solution or one tests the stability of homogeneous configurations against inhomogeneous perturbations.
	    While the minimization strategy obviously provides the full information about the thermodynamic ground states, it either requires a parametrization of the inhomogeneous condensate or one minimizes the condensate in a lattice discretized theory, which demands expensive numerical computations including continuum and infinite volume limit.
	    Especially in higher-dimensional models there are open questions about the preferred form and dimensionality of the inhomogeneous modulation.
	    The stability analysis method on the other hand requires no ansatz and is comparatively cheap and flexible, meaning that it can be used in a straightforward way with various regularization schemes.
	    However, it comes with the disadvantage that it does not give conclusive information about the preferred shape of the inhomogeneous condensate. 
	    Moreover, the existence of an instability of $\HomMin$ towards an inhomogeneous perturbation is a sufficient, but not necessary condition for the existence of an \gls{ip}.
	    
	    Stability analyses are our main tool to explore the existence and properties of \glspl{ip} in the present work.
	    This approach has been used extensively in investigations of the \gls{njl} model as well as other four-fermion and Yukawa models within various \glspl{rs}, e.g, in \Rcite{Koenigstein:2023yzv,Pannullo:2023cat,Pannullo:2023one,Pannullo:2021edr,Koenigstein:2021llr,Buballa:2020nsi,Buballa:2020xaa,Buballa:2018hux,Tripolt:2017zgc,Braun:2014fga,Nakano:2004cd,Braun:2014fga,Braun:2015fva,Koenigstein:2024cyq,Dobereiner:1989jb}.
	    Therefore, we only recapitulate the core idea and refer to \Rcite{Buballa:2018hux} for a discussion and derivation in the context of the \gls{njl} model, to \Rcite{Buballa:2020nsi} for a discussion of stability analysis in combination with a lattice regularization and to \Rcite{Koenigstein:2021llr} for a detailed investigation of the range of validity of this method.
	    The starting point of a stability analysis is to consider an infinitesimal inhomogeneous perturbation of arbitrary form $\delta \phi(\vec{x})$ of a homogeneous field configuration $\mstar$.
	    One then expands the effective action in powers of this perturbation, where the first order correction is proportional to the left-hand side of the gap equation \labelcref{eq:gap}, which vanishes, if one considers a minimum of the homogeneous effective potential, i.e., $\mstar=0$ or $\mstar=\HomMin$, as an expansion point.
	    The coefficients of the second order correction correspond to the curvature of the effective action in the direction of the respective inhomogeneous perturbation.
	    For a perturbation of momentum $\x{q}$ to the bosonic field $\phi$, where $\phi$ is either $\sigma$ or $\pi_i$, this is given by the bosonic $1$-particle irreducible two-point vertex function $\gtwo[\phi](\x{q})$, for simplicity denoted as bosonic two-point function in the following.\footnote{We note that the bosonic two-point function is by derivation diagonal in momentum space for theories with local \gls{ff} interaction channels, see, e.g., \Rcite{Buballa:2020nsi, Koenigstein:2021llr, Pannullo:2023one}.}
	    If one finds negative values of $\gtwo[\phi](\x{q})$ for any $\phi$ and $\x{q} \neq 0$, there exists an inhomogeneous field configuration, which is energetically favored over the homogeneous expansion point.
	    If this expansion point is the global homogeneous minimum $\HomMin$, a negative value of $\gtwo$ signals that the corresponding inhomogeneous configuration is favored over any homogeneous configuration.
	    To summarize, the check for an \gls{ip} at given $\left(\mu,T\right)$, requires two steps, $(1)$ the determination of $\HomMin{}(\mu,T)$ and $(2)$ the computation of $\gtwo[\phi](\x{q})$, where negative values indicate the presence of an \gls{ip}.
	    
	    For the \gls{pv} and \gls{sc} schemes, $\gtwo[\phi](\x{q})$ can be split up into two divergent contributions stemming from the fermionic loop integral: $\ellOne$, which does not depend on $\x{q}$ and also appears in the gap equation \eqref{eq:gap}, and $\ellTwo(\x{q})$, see \cref{eq:app:gamma2splitup}.
		For the lattice discretizations, this is not possible and the remaining loop integral $\ell_{3, \phi_i}(\x{q})$ depends on the chosen lattice discretization, compare \cref{eq:Appendix:SLACTracePropTemporalintegration} and \cref{eq:Appendix:HybridTracePropTemporalintegration} respectively.
	    Further details on the evaluation of $\gtwo[\phi](\x{q})$ can be found in \cref{app:formulas_stability} and on the regularization schemes in \cref{sec:regularization}.
	  
    \subsubsection*{Moat regime}
    
	    A moat regime is characterized by a non-trivial bosonic dispersion relation with a global minimum at a non-zero momentum.
	    In contrast to an \gls{ip}, the two-point function, which allows to read off the dispersion relation, does not have to exhibit a negative value at this minimum.
	    A simple criterion characterizing a moat regime is a negative sign of the wave-function renormalization $\zPhys[\phi] = z_\phi(\Mass = \HomMin)$, $\left(\phi = \sigma, \pi_i\right)$ \cite{Koenigstein:2021llr}, where
	    \begin{align}
	    z_\phi(\Mass) = \frac{\dr \gtwo[\phi](q)}{\mathrm{d} q^2} \Big|_{q=0}. \label{eq:zsmall}
	    \end{align} 
	    This criterion might fail to detect a moat regime, when $\gtwo$ has a global minimum at a finite $q$ while having a local minimum at $q=0$, i.e., the two minima are separated by a maximum and the curvature is positive at $q=0$.
	    Such a situation would also be a moat regime as the global minimum at finite $q$ is the physically defining feature, which just would be invisible to our simple criterion.
	    We are, however, not aware of any work observing such a scenario in a renormalizable model.\footnote{Later, we will discuss a case in the \gls{njl} model (which is non-renormalizable), where the two-point function shows exactly this behavior.
	    It is, however, in a parameter region, where the chemical potential is larger than the respective regulator $\Lambda$. This is clearly outside the range of validity of the model as there is no notion of a separation of scales and the regulator is no longer the largest energy scale in the system.}
	    
	    The formulas obtained for $z_\phi(\bar \sigma, 0)$ in the considered \glspl{rs} can be found in \cref{app:formula_z}.
	    
	\subsection{Regularization schemes \label{sec:regularization}}

		Both the calculation of the homogeneous effective potential and the stability analysis involve the evaluation of fermionic loop integrals that are \gls{uv}-divergent. 
		In this work, we use and compare several schemes and we discuss in the following how the schemes regulate such integrals.
		We consider the generic integral
		\begin{align}
			I=\tfrac{1}{\beta} \sum_{n=-\infty}^{\infty} \int \tfrac{\dr^3 p}{(2 \uppi)^3} f(\nu_n,\vecb{p},\Mass), \label{eq:reg_example}
		\end{align}
		where $\vecb p$ are spatial loop momenta, $\nu_n$ are the fermionic Matsubara frequencies and $f$ is the integrand, which is a combination of free fermion propagators with mass $\Mass$.
		This integral is \gls{uv}-divergent when $f\propto p^a$ with $a\geq-3$ for large $p$.
		This is the case for the integrals $\ellOne$ and $\ellTwo$ (for the \gls{sc} and \gls{pv} scheme) as well as $\ell_{3, \phi_i}$ (for the lattice discretizatons). 
		
	\subsubsection{Pauli-Villars}
	
		The \gls{pv} regularization scheme introduces additional copies of the integrand with differing coefficients and masses.
		These can be interpreted as contributions from both bosonic and fermionic replica fields of the original fermionic fields.
		The integral \labelcref{eq:reg_example} is then of the form
		\begin{align}
			I^{(\mathrm{\gls{pv}})}=\tfrac{1}{\beta} \sum_{n=-\infty}^{\infty} \int \tfrac{\dr^3 p}{(2 \uppi)^3}\pvsumck f(\nu_n,\vecb{p},\Mass_k), \label{eq:reg_example_PV}
		\end{align}
		where $\Mass_k=\sqrt{\Mass^2+\alpha_k \regulator}$ and the $k=0$ term with $\alpha_0=0$ is the unregularized contribution.
		The number of regulating terms $\NPV$ is not fixed and only bounded from below.
		For our investigation of the $(3+1)$-dimensional \gls{njl} model, $\NPV\geq2$ is necessary to render the most divergent integral finite. 
		The coefficients $c_k$ and $\alpha_k$ need to fulfill the conditions 
		\begin{align}
			\pvsumck={}0, \quad 
			\pvsumck \alpha_k={}0,
		\end{align}
		see, e.g., \Rcite{Klevansky:1992qe,Heinz:2015lua}.
		We follow existing literature and choose $\NPV=3$, $\vec{c}=(1,-3,3,-1)$ and $\vec{\alpha}=(0,1,2,3)$ \cite{Nickel:2009wj}.
	
	\subsubsection{Spatial momentum cutoff}
	
		With the \gls{sc} scheme one restricts the region of integration of the spatial loop momenta to a $3$-dimensional sphere of radius $\regulator$.
		The integral \labelcref{eq:reg_example} is then replaced by
		\begin{align}
			I^{(\mathrm{\gls{sc}})}=\tfrac{1}{\beta} \sum_{n=-\infty}^{\infty} \int \tfrac{\dr \Omega}{{(2 \uppi)^3}} \int_0^\regulator \dr p\,p^2\,f(\nu_n,\vecb{p},\Mass), \label{eq:reg_example_SC}
		\end{align}
		where $\int \dr \Omega$ represents the angular integration over the loop momenta.
		One of the main drawbacks of this scheme is that shifts of the integration variable $p$ lead to a modification of the integration boundaries, which can severely complicate the evaluation of such integrals.\footnote{A possible variation of such cutoff schemes is to introduce an energy cutoff instead of a momentum cutoff as proposed, e.g., in \Rcite{Adhikari:2016jzc,Pereira:2023rfr}.}
		
	\subsubsection{Lattice regularizations\label{sec:reg_lattice}}
	
		Lattice regularizations lead to momentum integrals similar to those in the \gls{sc} scheme.
		For infinite spacetime volume\footnote{For finite spacetime volume, one finds that integrals over momenta in directions with a finite extent are replaced by sums. However, since we consider lattice regularizations almost exclusively in an infinite spacetime volume, i.e., $T=0$ and $1/V=0$, we abstain from giving the respective formulas.} \cref{eq:reg_example} becomes
			\begin{align}
			I^{(\mathrm{Lattice})}={}&\prod_{i=0}^{3}  \Bigg[\int_{-\regulatorLFT}^{\regulatorLFT} \dr p_i\, \Bigg] f'(p_0,\vecb{p},\Mass), \label{eq:reg_example_Lattice}
		\end{align}
		where $\regulatorLFT=\uppi / a$ with lattice spacing $a$ and a modified integrand, $f'$ containing free fermion propagators for the respective lattice fermion discretization as discussed in the following paragraphs.

	\paragraph{SLAC fermions}
	
		In the SLAC discretization \cite{Drell:1976bq,Drell:1976mj} the derivative operator is constructed such that the dispersion relation of a massless fermion matches the linear continuum dispersion relation of a massless fermion within the first Brillouin zone.\footnote{This discretization is sometimes also called the plane wave expansion.}
		As a consequence, one finds that this discretization fully respects chiral symmetry and is free of doublers.
		This construction, however, necessitates a discontinuity of the dispersion relation at the edge of the Brillouin zone.
		While this causes severe problems in gauge theories \cite{Karsten:1979wh}, the SLAC discretization has been successfully applied in the investigation of \gls{ff} models \cite{Wellegehausen:2017goy,Lenz:2019qwu,Lenz:2020bxk,Lenz:2020cuv,Lenz:2021kzo}.
		
		Within this discretization, one finds for the fermion propagator
		\begin{align}
			\quarkPropagatorSLAC[\X{p}] = \frac{-\iu \gamma_{\mu} \fermionDispSLAC\left(\X{p}_\mu\right) - \gamma_{0} \mu + \Mass}{\sum_{i=0}^{3} \left( \fermionDispSLAC(\X{p}_i)-\kron{i}{0}\,\iu \mu \right)^2 + \Mass^2} \label{eq:5:SLACProp}
		\end{align}
		with dispersion relation	
		\begin{align}
			\fermionDispSLAC\left(\X{p}_\mu\right) = 2\regulatorLFT \left(\frac{\X{p}_\mu}{2 \regulatorLFT}  - \floor*{\frac{1}{2}+\frac{\X{p}_\mu}{2 \regulatorLFT} }  \right),
			\label{eq:5:SLACDisp}
		\end{align}
		where $\floor{\argPlaceholder}$ denotes the floor function.

	\paragraph{Hybrid fermions}
	\label{sec:hybridFermions}
	
		This discretization is a combination of the SLAC discretization for the temporal derivative in the action and a naive lattice discretization for the spatial derivatives.
		The naive discretization also respects chiral symmetry, but exhibits $2^d$ fermion doublers, where $d=3$ is the number of naively discretized dimensions.
		It gives rise to a sinusoidal dispersion relation and one finds for the fermion propagator
		\begin{align}
			&\quarkPropagatorHybrid[\X{p}] =\label{eq:5:HybridProp}\\
			={}& \frac{-\iu \gamma_{0} \left[\fermionDispSLAC\left(\X{p}_0\right) -\iu \mu \right] - \iu \sum_{i=1}^{3} \gamma_{i} \sin\left( a \x{p}_i \right) + \Mass}{\left( \fermionDispSLAC(\X{p}_0)-\iu \mu\right)^2 + \sum_{i=1}^{3} \sin^2\left(a\x{p}_i\right)+\Mass^2}. \nonumber
		\end{align}
	
		The existence of fermion doublers in this discretization necessitates additional modifications of the action.
		Without such modifications, one finds that at large momenta the bosonic fields mediate unphysical interactions between different doubler species in the Yukawa-interaction terms \cite{Cohen:1983nr}.
		One can suppress these unwanted interactions by introducing in momentum space a weight function $\tilde{W}$ to the Yukawa-interaction terms
		\begin{align}
			\tilde{\bar{\psi}}({p}) \, (\tilde{\sigma}&({p}-{q}) + \iu\gamma_5 \, \pmb{\tau} \,\tilde{\pmb{\pi}}({p}-{q}))\,  \tilde{\psi}({q})\quad  \\ \nonumber
			\to{}& \tilde{\bar{\psi}}({p}) \,\weightFunctionFourier[\mathrm{X}](\x{p}-\x{q}) (\tilde{\sigma}({p}-{q}) + \iu\gamma_5 \, \pmb{\tau} \,\tilde{\pmb{\pi}}({p}-{q}))\,  \tilde{\psi}({q}),
		\end{align}
		where $\tilde{\bar{\psi}},\tilde{\psi},\tilde{\sigma},\tilde{\pmb{\pi}}$ are the Fourier components of the respective fields and $\weightFunctionFourier[\mathrm{X}](\x{p}-\x{q})= \prod_{i=1}^{3} \weightFunctionFourierOneD[\mathrm{X}](p_i-q_i)$.
		There are infinitely many possible choices for $\weightFunctionFourierOneD[\mathrm{X}]$, which lead to a suppression of the interactions between the doublers and to identical results for renormalizable theories in the continuum limit.
		We consider the two possibilities	
			\begin{align}
				\weightFunctionFourierCosOneD(p_\mu)=\frac{1+\cos(p_\mu)}{2}
				\label{eq:cosWeighting1D}
			\end{align}
		and
			\begin{align}
				\weightFunctionFourierHeavisideOneD(p_\mu)=\Theta\left(1-\Abs{p_\mu} \tfrac{2a}{\uppi}\right).
			\label{eq:heavisideWeighting1D}
			\end{align}
		Since the $3+1$-dimensional \gls{njl} model is a theory, where one cannot take the continuum limit, these two choices are not equivalent, and we thus consider them as separate \glspl{rs} denoted as \hybridCos{} and as \hybridTheta{}, respectively.

	\subsubsection{No medium regularization}
	\label{sec:NMR_theory}
	
		Within some \glspl{rs} one can separate the integral $I$ into a medium part dependent on $\mu$ and $T$ and a vacuum contribution independent of such thermodynamical quantities. 
	    In this work, this is the case for the \gls{pv} scheme and \gls{sc} scheme.
		In our context, only the vacuum contributions are \gls{uv}-divergent while medium contributions stay finite, since the respective integrand vanishes at large momenta due to exponential suppression by the Fermi-Dirac distribution functions. 
		After regularization and for chemical potentials and/or temperatures that are in the order of or larger than the regulator, one finds that the fall-off of the integrand in the medium contribution is modified, e.g., cut off in the case of the \gls{sc} scheme.
		In a renormalizable theory, where the regulator is removed at the end, it does not make a difference, whether the whole theory or only the vacuum parts are regulated.
		However, as we need to keep the regulator finite it might be advantageous to reduce the regularization artifacts by only regulating the \gls{uv}-divergent parts.
		We indicate such a  version of a \gls{rs}, that allows the separation of vacuum and medium parts, as \gls{nmr}.
		
		Applying \gls{nmr} was found to have a significant positive impact on the homogeneous phase diagram \cite{Kohyama:2015hix} and physical quantities such as the speed of sound, which can violate causality in a large region of the phase diagram for the \gls{sc} without \gls{nmr} \cite{Pasqualotto:2023hho}. 
		
		In our case, \gls{nmr} is equivalent to renormalization group consistency, which is discussed in \Rcite{Braun:2018svj} and provides a systematic approach to correct for regularization artifacts in the medium contributions without explicitly separating them from the vacuum parts.

	\subsection{Parameter tuning}
	\label{sec:parameterfitting}
	
		As discussed in \cref{sec:hompot}, we need to consider the regulator $\regulator$ as a parameter of the theory, in addition to the coupling $G$.
		When interpreting the \gls{njl} model as a low-energy model for chiral symmetry breaking in \gls{qcd}, the value of $\Lambda$ represents the energy scale, where the gluon dynamics start to become dominant and the approximation of the gluon-mediated interactions as \gls{ff} vertices is no longer valid \cite{Klevansky:1992qe}.
		
		The coupling $G$ and the regulator $\Lambda$ can be tuned such that certain standard observables assume physically motivated values.
		We follow an approach proposed in \Rcite{Klevansky:1992qe}, which has become a standard in \gls{njl} model investigations (see, e.g., \Rcite{Buballa:2003qv, Hands:2004uv, Buballa:2014tba, Buballa:2018hux, Pasqualotto:2023hho}), namely to fix the pion decay constant $f_\pi= 88\, \mathrm{MeV}$.\footnote{In \gls{qcd} with physical quark masses, one would expect $f_\pi\approx 92.4 \, \mathrm{MeV}$ \cite{ParticleDataGroup:2022pth}. However, using chiral perturbation theory one finds for vanishing bare quark masses $f_\pi\approx88\, \mathrm{MeV}$ \cite{Gasser:1984gg,Gerber:1988tt}.}
		The constituent quark mass in the vacuum $M_0 = \HomMin(\mu=0, T = 0)$ is the second key quantity that is used as input in the tuning. 
	    To relate $f_\pi$ and $M_0$ to $G$ and $\regulator$, we follow \Rcite{Klevansky:1992qe} and present corresponding formulas for $G$ and $f_\pi / M_0$ in \cref{app:formulas_parameterfitting}. 
				
	\section{Regularization scheme dependence of the inhomogeneous phase at zero temperature}
	\label{sec:IPResults}
	
	\subsection{General phase structure}
	\label{sec:general_phase_structure}
	
\begin{figure*}
			\centering
			\includegraphics[width=\linewidth]{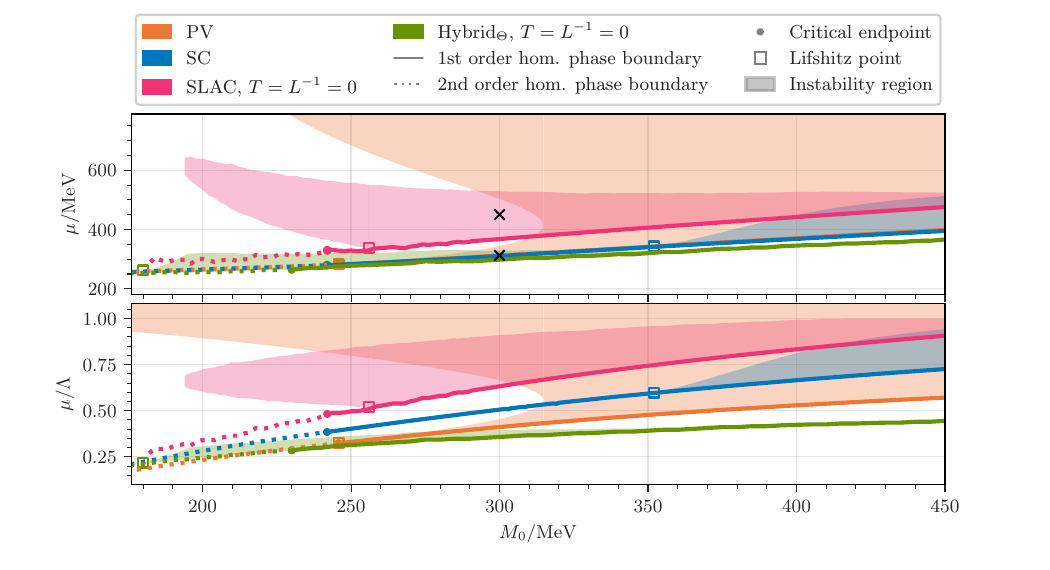}
			\caption{The homogeneous phase diagram and the regions of instability towards an \gls{ip} in the $(\Mzero,\mu)$-plane at $T=0$ within four \glspl{rs}. Both plots depict the same data, where the upper plot the chemical potential is given in units of MeV, while the lower plot gives $\mu$ in units of the respective $\Lambda$, which is a function of $\Mzero$. The solid and dotted lines represent first and second order homogeneous phase boundaries respectively obtained by a minimization of the effective potential.
			The circle represents the \gls{cep} that separates these boundaries.
			The shaded regions are the instability regions where the bosonic two-point function exhibits negative values for non-zero spatial momenta.
			The open squares indicate the position of the \gls{lp}.
			The black crosses mark $(\Mzero,\mu)$ values chosen for the plots in \cref{fig:twpsmzero300}.
			The data for the \hybridCos{} scheme is not shown as it does not exhibit any instability region and the homogenoeus phase boundary is identical to that obtained in the \hybridTheta{} scheme.}
			\label{fig:combi}
		\end{figure*}

		We start the discussion by considering the central result of this work: the phase structure within different \glspl{rs} in the $(\Mzero,\mu)$-plane at $T=0$ as shown in \cref{fig:combi}.
		From this figure it is obvious that there is a strong dependence of both the existence and the extent of \glspl{ip} on the \gls{rs} and the regulator $\regulator$.
		This ambiguity indicates that predictions of an \gls{ip} in the \gls{njl} model have rather limited implications for the phase structure of \gls{qcd}.
		 
		In \cref{fig:combi}, moving along the horizontal axis corresponds to considering different values for $\Mzero$, i.e., different parameter sets with $\piondecay$ being fixed. 
		It is noteworthy that the value of $\piondecay$ only sets the scale in $\mathrm{MeV}$ in the plots and does not change the phase structure, see \cref{sec:parameterfitting}.
		In the plot, the \gls{hbp} is located at small chemical potential, i.e., below the phase boundary lines.
		While the homogeneous phase boundaries (solid and dottes lines) are qualitatively similar between the \glspl{rs}, we find that the shape and extent of the \gls{ip} (shaded regions) in this plane is highly dependent on the choice of the \gls{rs}. 
		In particular, we note that there is not a single point in the $\left(\Mzero,\mu\right)$-plane, where all \glspl{rs} exhibit an \gls{ip} simulatenously.
		For small values of $\Mzero$, which correspond to large values of $\regulator$, we find that the \gls{ip} is absent within all \glspl{rs}, which is the only region where the \glspl{rs} agree with respect to the existence of the \gls{ip}.
			
		We note that the observed strong dependence of the existence and extent of the \gls{ip} in the \gls{njl} model on the \gls{rs} and the value of the regulator is not surprising:
		\Glspl{ip} appear for values of the chemical potential, which are of the order of the regulator, which facilitates strong regularization artifacts.
	
		In the following we briefly discuss the results obtained within each of the \glspl{rs}:
		\begin{itemize}
			\item The \gls{pv} scheme leads to the largest \gls{ip} in the $(\Mzero,\mu)$-plane.
			One finds that the \gls{cep} and the \gls{lp} coincide in agreement with previous studies \cite{Nickel:2009wj,Carignano:2014jla}. 
			For intermediate values of $\Mzero<315 \mathrm{MeV}$, one finds two separated regions: the ``inhomogeneous island'', which is connected to the \gls{hbp} and the ``inhomogeneous continent'' \cite{Carignano:2011gr}, which is located at larger values of $\mu$ and disconnected from the \gls{hbp}.
			For small values of $\Mzero$, the ``inhomogeneous island'' is absent and only the ``inhomogeneous continent'' is present.
			\item Within the \gls{sc} scheme the \gls{ip} in the $(\Mzero,\mu)$-plane is significantly smaller compared to that of the \gls{pv} scheme.
	        One finds that the \gls{cep} and the \gls{lp} are separated, which is due to surface terms that arise in the underlying loop integrals \cite{Carignano:2014jla}.
	        \item Within the SLAC discretization one finds that an ``inhomogeneous continent'' splits off at an intermediate value of $\Mzero$ similar as in the \gls{pv} scheme.      
	        Moreover, a splitting of the \gls{cep} and the \gls{lp} is observed similar to the findings within the \gls{sc} scheme. 
	        The extent of the \gls{ip}, however, is larger than that of the \gls{sc} scheme.
	        The depicted phase boundaries exhibit small but clearly visible statistical fluctuations, which are a consequence of the stochastic evaluation of the underlying momentum integrals.
	        \item Within the \hybridTheta{} scheme, an \gls{ip} is present for small values of $\Mzero$ and $\mu$. The \gls{cep} and \gls{lp} are separated, but the splitting is in the opposite direction compared to the \gls{sc} and SLAC scheme, i.e., the \gls{lp} is located at larger values of $\Mzero$ than the \gls{cep}. This is mostly likely due to the effective vector interactions that arise as discretization artifacts in Yukawa-theories with schemes that exhibit fermion doublers such as the Hybrid schemes \cite{Cohen:1983nr,Lenz:2020bxk}.
	        Such vector actions are known to split the \gls{cep} and the \gls{lp} in the \gls{njl} model such that the \gls{lp} is located at a higher temperature \cite{Carignano:2010ac,Carignano:2018hvn}.
	        
	        \item It is particularly important to note that no instability region is observed at all in the \hybridCos{} scheme. Since the homogeneous phase boundary line is by definition identical to the one of the \hybridTheta{} scheme, we abstain from plotting it separately.
		\end{itemize}
	    In the following sections, we will elaborate more on the significant, qualitative differences in the results obtained within different \glspl{rs}.
		
		A related finding with respect to the \gls{rs} scheme dependence of the \glspl{ip} is the study of the phase diagram of the $d+1$-dimensional \gls{gn} model in \Rcite{Pannullo:2023cat,Koenigstein:2023yzv, Pannullo:2024hqj}, where an \gls{ip} is present for $d \in [1,2)$ and absent for $d \in [2,3)$.
		The \gls{gn} model is also non-renormalizable for $d=3$ and the results from \Rcite{Pannullo:2023cat,Koenigstein:2023yzv} can be interpreted as a study of the $3+1$-dimensional \gls{gn} model using dimensional regularization.
		A small $d$ would then correspond to a strongly-regularized theory, i.e., a small value of $\regulator$ within \glspl{rs} considered in this work.
	 	The integral expression for the bosonic two-point function of the $\sigma$ field is mathematically identical in the \gls{gn} model and the \gls{njl} model (independent of $d=1,2,3$) and the \gls{ip} in the \gls{njl} model typically is realized through inhomogeneities in the $\sigma$ condensate \cite{Nickel:2009wj}.
		Thus, these studies provide results within a sixth \gls{rs} scheme pointing towards the non-existence of an \gls{ip} for higher values of $\regulator$ (high values of $d$ in dimensional regularization), consistent with the other five \glspl{rs}. 

	\subsection{Two-point functions}
	\label{sec:twps}
	
		\begin{figure*}
			\centering
			\includegraphics[width=\linewidth]{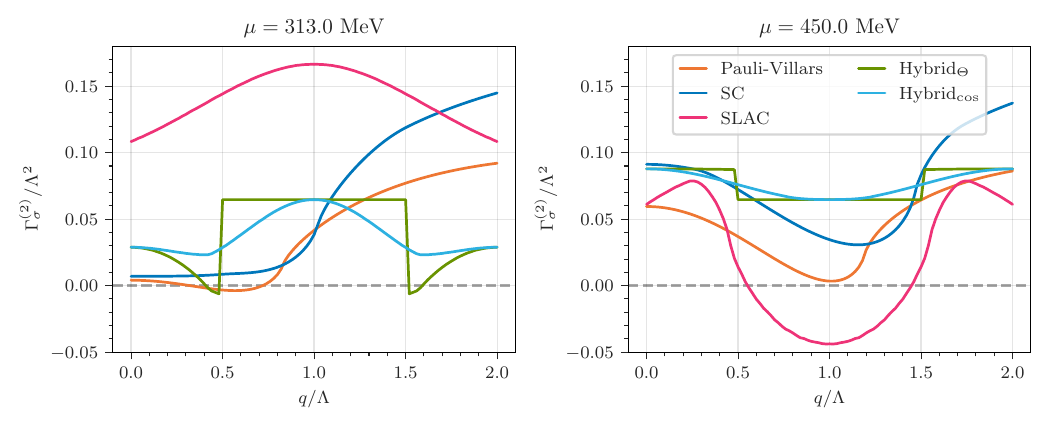}
			\caption{The bosonic two-point function $\gtwo[\sigma]$ as a function of the bosonic momentum $q$ for chemical potential ${\mu=313\, \mathrm{MeV}, \, 450 \, \mathrm{MeV}}$ at $T=0$ and $\Mzero=300\, \mathrm{MeV}$. Different colors represent different \glspl{rs}.
			The black crosses in \cref{fig:combi} mark points in the $(\Mzero,\mu)$-plane these plots correspond to.}
			\label{fig:twpsmzero300}
		\end{figure*}
		In this section, we study the rather different behavior of the two-point functions within all five \glspl{rs} to illustrate possible reasons for the vastly different phase diagrams discussed in the previous section and shown in \cref{fig:combi}.	
		\cref{fig:twpsmzero300} depicts the bosonic two-point function $\gtwo[\sigma]$ as a function of the bosonic momentum $q$ for all considered \glspl{rs} and selected values of the chemical potential $\mu$, $T=0$, ${\Mzero=300\, \mathrm{MeV}}$.
	
		One finds that the \gls{pv} and \gls{sc} scheme exhibit a crudely similar behavior of the two-point function.
		There are, however, visible differences in terms of a vertical shift as well as of the depth of the minima, which cause the absence of an instability for the \gls{sc} scheme, while it is present in the \gls{pv} scheme (see the left plot).
	
		The lattice discretizations exhibit a periodic two-point function, which is a consequence of the periodic dispersion relations on the lattice.
		The SLAC discretization exhibits a non-zero positive slope at $q=0$ for all $\mu$, which is a consequence of the jump in the fermionic dispersion relation at the edge of the Brillouin zone.
		This can lead to a maximum in the two-point function at a finite momentum, while a minimum can be present for even larger momenta, as is the case in the right plot of \cref{fig:twpsmzero300}. 
	
		The Hybrid discretizations exhibit rather different two-point functions because of differences in the respective weighting functions.
		This is especially evident with the \hybridTheta{} scheme, where the two-point function jumps at $q=\regulatorLFT/2$ to large positive values due to the discontinuos weighting function (cf.~\cref{eq:heavisideWeighting1D}).
		At the chosen chemical potential in the left plot of \cref{fig:twpsmzero300}, the two-point function is negative for small momenta signaling an instability towards an \gls{ip}.
		For increasing $\mu$ these negative intervals are shifted towards larger $q$ and finally disappear, because they are suppressed by the weighting function.	
		In contrast, the smooth weighting function in the \hybridCos{} scheme (cf.~\cref{eq:cosWeighting1D}) suppresses the two-point function already for small momenta causing the absence of instabilities which are present in the \hybridTheta{} scheme.
		
	\subsection{Characteristic momentum}

		The characteristic momentum $\qChar$ is defined as that momentum, where the bosonic two-point function $\gtwo[\sigma](q)$ assumes its minimal value,
		\begin{align}
			\qChar = \argmin_{q} \gtwo[\sigma](q).
		\end{align}
		Even though this is not necessarily the dominating momentum of the inhomogeneous field configuration, which can, in principle, be obtained by minimizing $\seff$, it is expected to be similar and, thus, a useful quantity to characterize an \gls{ip}.
		On a second-order phase boundary between an \gls{sp} and an \gls{ip}, $Q$ corresponds to the momentum of the inhomogeneous condensate \cite{Koenigstein:2021llr}.
		By considering energy gaps at the Fermi surface and following the argumentation of the Peierls stability, one expects that $\qChar\approx 2\mu$ (see, e.g., \Rcite{Kojo:2009ha, Buballa:2014tba} for a discussion in the context of these models).
		However, the exemplary two-point functions discussed in \cref{sec:twps} have shown that this is in general not the case for the lattice regularization schemes.
		
\begin{figure*}
			\centering
			\includegraphics[width=\linewidth]{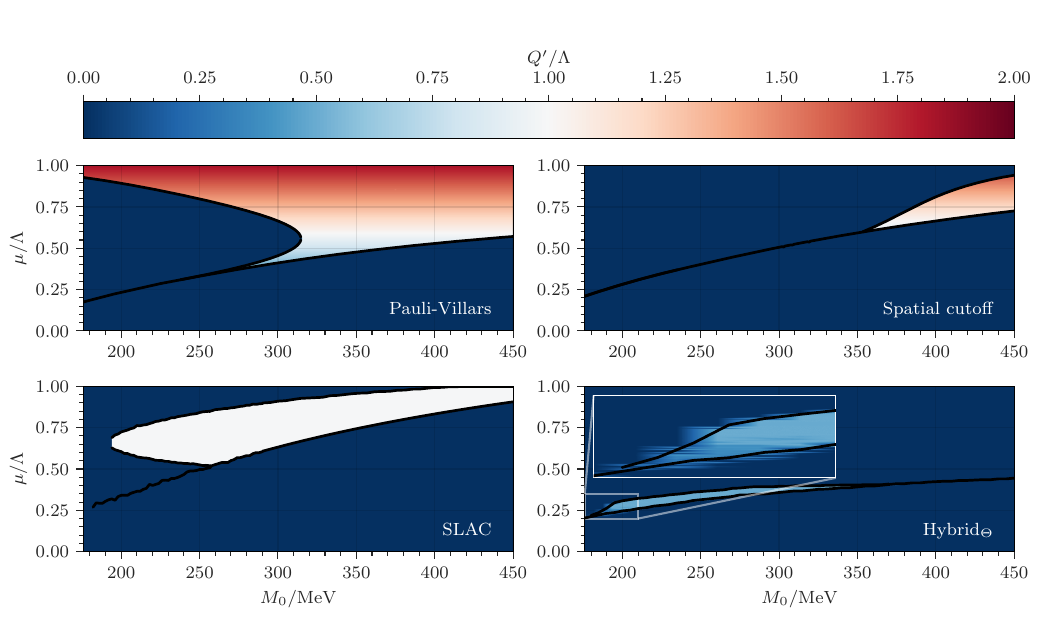}
			\caption{
				The characteristic momentum $Q'(\mu,T)$ in the $(\Mzero,\mu)$-plane at $T=0$ within four \glspl{rs}.
			}
			\label{fig:qscan4pane}
		\end{figure*}

		To investigate the behavior of the characteristic momentum systematically, we define 
		\begin{align}
			Q'(\mu,T) = \begin{cases}
				\qChar(\mu,T) & \text{if } \gtwo(\qChar) < 0\\
				0 & \text{otherwise}
			\end{cases},
		\end{align}
		because only a characteristic momentum associated to an instability is relevant for this discussion.
		In \cref{fig:qscan4pane}, $Q'(\mu,T)$ is shown within different \glspl{rs} in the $\left(\Mzero,\mu\right)$-plane at $T=0$. 
		Within the \gls{pv} and \gls{sc} schemes one finds that the characteristic momentum indeed follows the expectation of $\qChar\approx 2\mu$.
		While this is the expected behavior, it is important to note that consequently $\qChar\gtrsim\regulator$ within most parts of the instability region.
		This is highly problematic, since it prevents separation of scales in the system and the regulator $\Lambda$ does not serve as the largest energy scale in the system anymore.
		In contrast, within the SLAC discretization one finds that the characteristic momentum is constant, $\qChar=\regulatorLFT$ within the instability region ($\regulatorLFT$ is the maximal momentum that is available on the lattice, as defined in \cref{sec:reg_lattice}).
		With the \hybridTheta{} scheme, one finds that the characteristic momentum is almost constant $\qChar=\regulatorLFT/2$ within the instability region. 
		This corresponds to the threshold, when the weighting function sets in.
		Note that in the two latter cases $Q$ is not related to the chemical potential $\mu$, as expected according to the discussion above, but to the momentum cutoff $\regulatorLFT \propto 1 / a$.
		A corresponding \gls{ip} is then clearly an artifact of the regularization.
		
	\subsubsection{Dominating versus characteristic momentum within the	SLAC discretization}
	
		As previously discussed, the characteristic momentum $\qChar$ might not be identical to the dominating momentum of the corresponding inhomogeneous field configuration.
		We, therefore, investigate such field configurations for the SLAC discretization to check whether the inhomogeneous condensate actually oscillates with $Q=\regulatorLFT$.
		To this end, we consider the SLAC discretized action in a finite spacetime volume $T^{-1} \times L^3= 7.7 \times (7.4)^3 \,\mathrm{MeV}^{-4}= 30 \times 29^3\, a^4$.
		Then, we perform minimizations of the effective action with respect to the bosonic fields.
		We restrict the minimization to $1$-dimensional field configurations, to limit the numerical effort.
		This assumption is based on previous findings \cite{Carignano:2012sx} where $1$-dimensional field configurations were mostly favored.
		No further assumptions about the field configurations are made.
		
\begin{figure}
			\centering
			\includegraphics{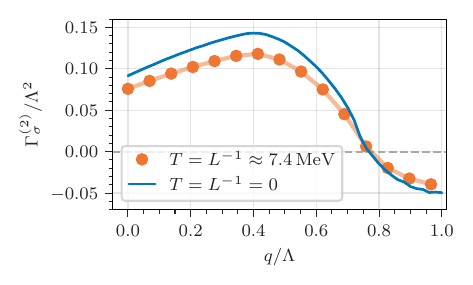}
			\caption{
				The bosonic two-point function $\gtwo[\sigma]$ as a function of the bosonic momentum $q$ for the SLAC discretization in infinite $L^{-1}=T=0$ and finite spacetime volume $T^{-1} \times L^3= 7.7 \times (7.4)^3 \,\mathrm{MeV}^{-4}= 30 \times 29^3\, a^4$ at $\Mzero=287\,\mathrm{MeV}$ and $\mu = 380\, \mathrm{MeV}$.
			}
			\label{fig:slac1dgamma288}
		\end{figure}

\begin{figure}
			\centering
			\includegraphics{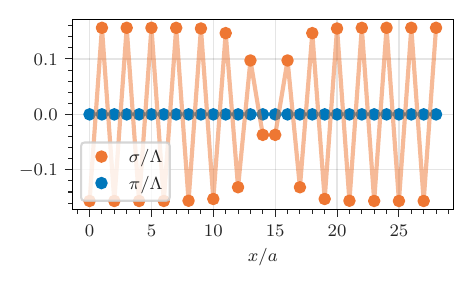}\\[3mm]
			\includegraphics{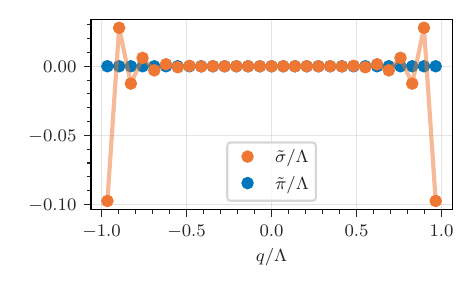}
			\caption{
				The energetically preferred  bosonic field configuration in position \textbf{(top)} and momentum space \textbf{(bottom)}  for the SLAC discretization in finite spacetime volume $T^{-1} \times L^3= 7.7 \times (7.4)^3 \,\mathrm{MeV}^{-4}= 30 \times 29^3\, a^4$ at $\Mzero=287\,\mathrm{MeV}$ and $\mu = 380\, \mathrm{MeV}$.
			}
			\label{fig:slac1dpos88}
		\end{figure}

		In a first step, we check and confirm that the two-point functions $\gtwo[\sigma](q)$ in the infinite spacetime volume and in the finite spacetime volume exhibit a similar behavior for $\Mzero=287\,\mathrm{MeV}$ and $\mu = 380\, \mathrm{MeV}$ (see \cref{fig:slac1dgamma288}).
		The upper plot in \cref{fig:slac1dpos88} then depicts the energetically preferred field configuration which was determined via the aforementioned minimization.
		We find $\pi(x)=0$, which is in agreement with other continuum findings \cite{Heinz:2015lua}, while the $\sigma$ field oscillates with the largest possible momentum, which is in agreement with the expectation from the characteristic momentum.
		This is also shown by the momentum distribution in the lower plot in  \cref{fig:slac1dpos88}. 
		This confirms that the \gls{ip} within the SLAC discretization indeed prefers a condensate which oscillates with the maximally possible momentum $\regulatorLFT = \pi / a$.
		As discussed above, this is clearly an artificial behavior related to the \gls{rs} but not a physically meaningful \gls{ip}.
		
	\subsection{No medium regularization}
	
		In the above sections, we presented evidence and argued that strong regularization artifacts are present for $\mu/\regulator \gtrsim 1$, which is the relevant region for \glspl{ip} in the \gls{njl} model. 
		As discussed in \cref{sec:NMR_theory}, \gls{nmr} is a procedure to remove regularization artifacts in the medium contributions of a given \gls{rs}.
		Thus, it is expected that this procedure reduces the observed discrepancies between the different \glspl{rs}.
		This is studied in the following within the \gls{sc} and \gls{pv} schemes.
		
\begin{figure*}
			\centering
			\includegraphics{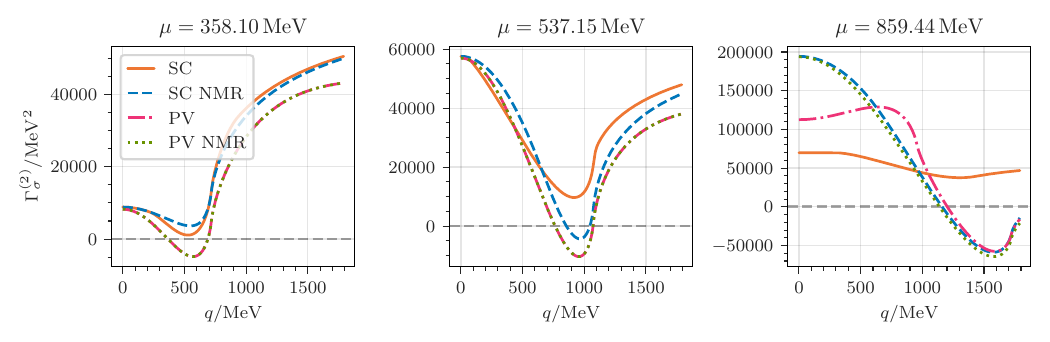}
			\caption{
				The bosonic two-point function $\gtwo[\sigma]$ as a function of the bosonic momentum $q$ for three different values of the chemical potential $\mu$, $T=0$ and $\Mzero=350\, \mathrm{MeV}$ within the \gls{pv} and the \gls{sc} scheme with and without \gls{nmr}.
				For $\mu<\regulatorPV$ (as it is the case in the left and middle plot), the two-point function with the \gls{pv} scheme is unaffected by the \gls{nmr} procedure.
				The black crosses in \cref{fig:combirgconsistent} mark the points in the $(\Mzero,\mu)$-plane these plots correspond to (the position of the most right plot in the  $(\Mzero,\mu)$-plane is outside the plot range).
			}
			\label{fig:nmrgammatwocomparisonsamemumev}
		\end{figure*}

		\cref{fig:nmrgammatwocomparisonsamemumev} depicts the bosonic two-point function $\gtwo[\sigma]$ as a function of the bosonic momentum $q$ for three values of the chemical potential $\mu$ at $T=0$ with the \gls{pv} and the \gls{sc} scheme with and without \gls{nmr}.
		One finds that the \gls{nmr} has no impact on the results with the \gls{pv} scheme for $\mu<\regulatorPV$ ( $\regulatorPV$ corresponds to the smallest mass of the regulating terms), which is a manifestation of the silver blaze phenomenon \cite{Cohen:2003kd}.
		Similarly, the results of the \gls{sc} scheme change marginally for small values of $\mu$, but the impact of \gls{nmr} becomes more significant for larger values of $\mu$.
		At such values for $\mu$, we observe for both schemes an unexpected behavior of the two-point function for small momenta without \gls{nmr}, which is distinctly different from two-point functions in the instability regions as found in renormalizable $1+1$-dimensional models \cite{Braun:2014fga,Koenigstein:2021llr}.
		As visible in the right plot of \cref{fig:nmrgammatwocomparisonsamemumev}, the \gls{sc} scheme leads to a constant two-point function in an interval of small $q$, while within the \gls{pv} scheme one finds a positive curvature of the two-point function at $q=0$. 
		However, both schemes exhibit a minimum of the two-point function for large momenta.
		We interpret this as artifacts stemming from the \glspl{rs}.
		These artifacts are removed when using the \gls{nmr}.
		The two-point functions then exhibit a qualitatively similar behavior for both schemes consistent with expectations from renormalizable $1+1$-dimensional models.
	
\begin{figure*}
			\centering
			\includegraphics[width=\linewidth]{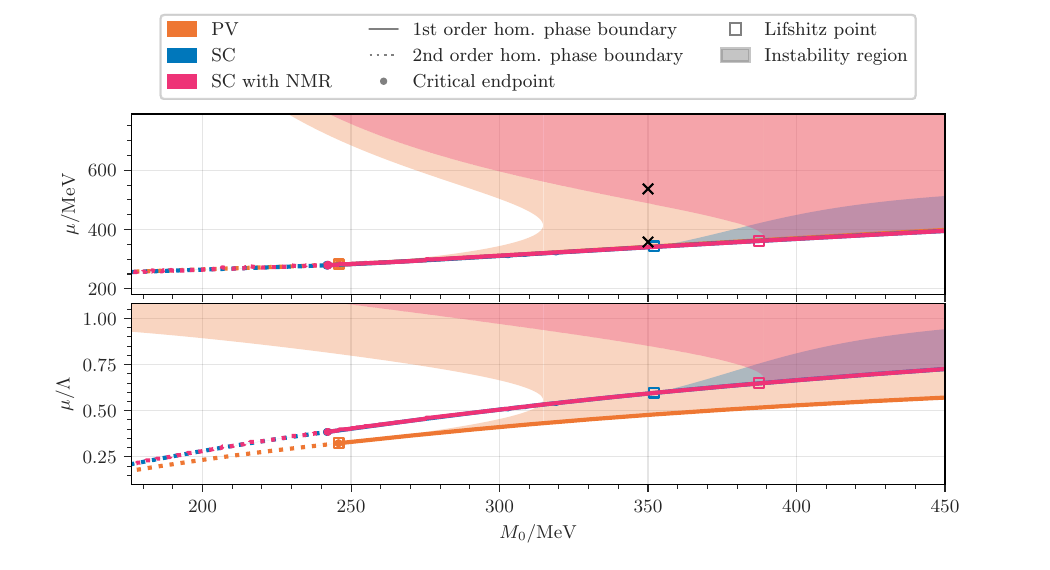}
			\caption{
				The homogeneous phase diagram and the regions of instability towards an \gls{ip} in the $(\Mzero,\mu)$-plane at $T=0$ for the \gls{pv} and the \gls{sc} scheme with and without \gls{nmr}.
				Both plots depict the same data, but in the upper plot the chemical potential is given in units of MeV, while the lower plot gives $\mu$ in units of the respective $\Lambda$, which is a function of $\Mzero$. The solid and dotted lines represent first and second order homogeneous phase boundaries respectively obtained by a minimization of the effective potential.
				The circle represents the \gls{cep} that separates these boundaries.
				The shaded regions are the instability regions where the bosonic two-point function exhibits negative values for non-zero spatial momenta.
				The open squares indicate the position of the \gls{lp}.
				The black crosses mark $(\Mzero,\mu)$-values chosen for the plots in \cref{fig:nmrgammatwocomparisonsamemumev} (the position of the most right plot in the  $(\Mzero,\mu)$-plane is outside the plot range).
			}
			\label{fig:combirgconsistent}
		\end{figure*}

		\cref{fig:combirgconsistent} depicts the effect of \gls{nmr} on the \gls{ip} in the $(\Mzero,\mu)$-plane at $T=0$ for the \gls{pv} and the \gls{sc} scheme.
		The \gls{ip} is unaffected for the \gls{pv} scheme in the depicted region due to the mentioned silver-blaze like behavior and, thus, we only plotted the results without \gls{nmr}.
		For the \gls{sc} scheme, one finds that the \gls{ip} has a different shape and a significantly larger extent.
		Interestingly, there is a region where the \gls{sc} scheme without \gls{nmr} exhibits an instability region, while the \gls{sc} scheme with \gls{nmr} does not.
		We recognize that the shape of the \gls{ip} in the \gls{sc} scheme with \gls{nmr} looks qualitatively similar to the shape of the \gls{ip} in the \gls{pv} scheme without \gls{nmr} besides a global shift in $\Mzero$.
		The residual differences are caused by differences in the vacuum parts of the relevant quantities, where \gls{nmr} has no effect.
		Overall, our results suggest that the \gls{nmr} procedure effectively reduces \gls{rs} artifacts in the two-point functions, which leads to more consistent \glspl{ip} between the \gls{pv} and the \gls{sc} schemes.
		However, even when using \gls{nmr}, it is still unclear whether the \gls{njl} model has any predictive power for values of $\mu$ of the order or larger than the cutoff, where an \gls{ip} can be observed.


	\section{Regularization scheme (in)dependence of the moat regime}
	\label{sec:moatres}
	
		Following our analysis of the \gls{ip}, we turn to the study of the moat regime.
		We focus on the \gls{pv} and the \gls{sc} schemes as these schemes give better insight into the analytic structure of the central quantities to understand this phenomenon.
		Both schemes exhibit significantly different results for the extent and shape of the \gls{ip} in the $(\mu, \Mzero)$-plane at zero temperature (see \cref{sec:IPResults}) and thus an agreement of their moat regimes is not guaranteed.
		Moreover, while the SLAC discretization experiences a moat regime (see, e.g., right plot in \cref{fig:twpsmzero300}), the non-zero positive slope of the bosonic two-point function at $q=0$ causes our indicator $\zPhys[\sigma]$ to be undefined.\footnote{The indicator $\zPhys$ is the curvature of the bosonic two-point function at $q=0$ (see \cref{eq:zsmall}), which is an even function in $q$. Thus, a non-zero positive slope causes a diverging curvature at $q=0$.}
		 
		In this section we exclusively study $\zPhys[\sigma]$ as an indicator for the moat regime.
		This is sufficient as for all relevant parameters within the \gls{pv} and \gls{sc} scheme, either $\z[\sigma] \leq \z[\pi_i] \leq 0$ or $\z[\sigma],\z[\pi_i] \geq 0$ , i.e., in the presence of a moat regime the wave-function renormalization of $\sigma$ is the stronger indicator and otherwise both are positive.\footnote{This can -- in parts -- be inferred from considering the respective formulae \cref{eq:app:zPVGeneral,eq:app:zPVTZero,eq:app:zPVMZero,eq:app:zPVMTZero,eq:app:z3DCutoffGeneral,eq:app:z3DMZero} and was also confirmed numerically.}

	\subsection{The moat regime at zero temperature\label{sec:moat_zerot}}
	
\begin{figure*}
			\centering
			\includegraphics[width=\linewidth]{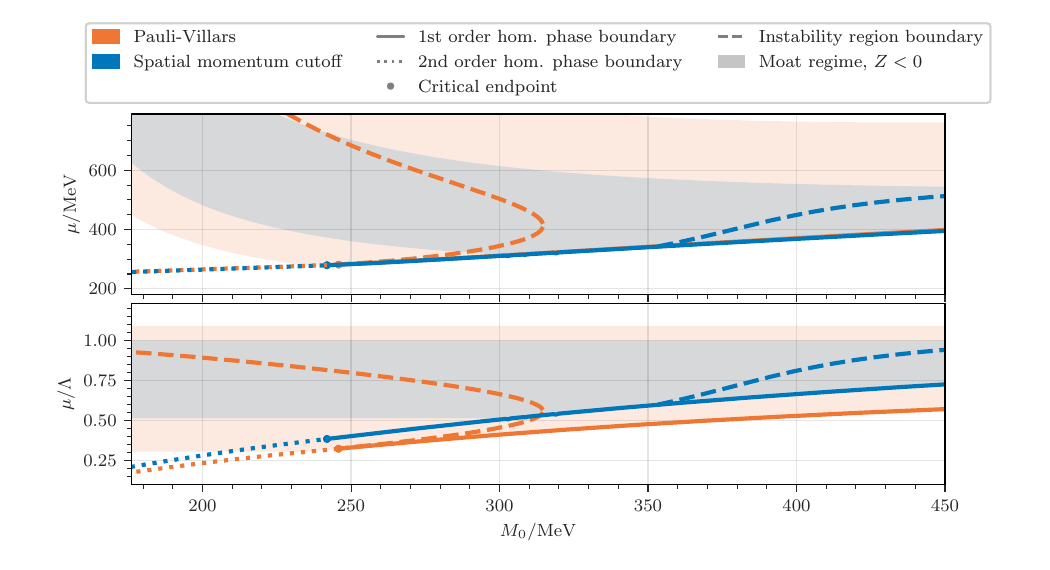}
			\caption{Homogeneous phase boundaries, instability region boundary and the moat regime from the stability analysis at $T=0$ in the $\left(\Mzero, \mu\right)$-plane for the \gls{pv} and the \gls{sc} regularization.
			Both plots depict the same data, but in the upper plot the chemical potential is given in units of MeV, while the lower plot gives $\mu$ in units of the respective $\Lambda$, which is a function of $\Mzero$. The solid and dotted lines represent first and second order homogeneous phase boundaries respectively obtained by a minimization of the effective potential.
			The circle represents the \gls{cep} that separates these boundaries.
			The dashed lines delimit the instability regions where the bosonic two-point function exhibits negative values for non-zero spatial momenta.
			The shaded regions depict the moat regime, where the wave-function renormalization $\zPhys[\sigma]$ is negative.}
			\label{fig:moattzero}
		\end{figure*}

		In \cref{fig:moattzero} we plot the moat regime as characterized by $\zPhys[\sigma] < 0$ in the $(\mu, \Mzero)$-plane at zero temperature.
		As in \cref{fig:combi}, the upper and the lower plot show the same data and differ only in the units used to express the chemical potential (upper in units of $\mathrm{MeV}$, lower in units of the respective regulator value $\Lambda$). 
		As for the \glspl{ip}, the relevant values of the chemical potential are of the order of the regulator. 
		For both the \gls{pv} and the \gls{sc} scheme, a moat regime is found in the \gls{sp} adjacent to the homogeneous first order phase transition (solid line) in a large range for $\Mzero$ (roughly $250 \, \mathrm{MeV}$ to $450\, \mathrm{MeV}$ for the \gls{pv} scheme and $300 \, \mathrm{MeV}$ to $450\, \mathrm{MeV}$ for the \gls{sc} scheme).
		For small values of $\Mzero$ the region of negative $\zPhys[\sigma]$ starts to disconnect from the \gls{hbp}.
	 	The shape of the moat regime in the $(\mu, \Mzero)$-plane is rather similar in both cases despite the large chemical potential (in units of the respective regulator).
	    In contrast, the instability region boundaries (dashed lines in \cref{fig:moattzero}) representing the \gls{ip} in both schemes are of completely different shape.
		This suggests that the moat regime might be a more robust feature of the \gls{njl} phase diagram in the $(\mu, \Mzero)$-plane when varying the respective regulator value and changing the \gls{rs}.
		Only for values of $\Mzero$, which are smaller than $200\, \mathrm{MeV}$, i.e., values significantly below those suggested by \gls{qcd}, one does not obtain negative $\zPhys[\sigma]$ for moderate values of $\mu\approx 300\,\mathrm{MeV}$.
		For these small values of $\Mzero$, however, one finds a moat regime at larger values of $\mu$, disconnected from the \gls{hbp}, which is most likely a regulator artifact.\footnote{This is similar to the disconnected inhomogeneous continent, that was identified as a likely regularization artifact \cite{Carignano:2011gr}.}

	\subsection{The moat regime at finite temperature}
	
\begin{figure*}
			\centering
			\includegraphics[width=\linewidth]{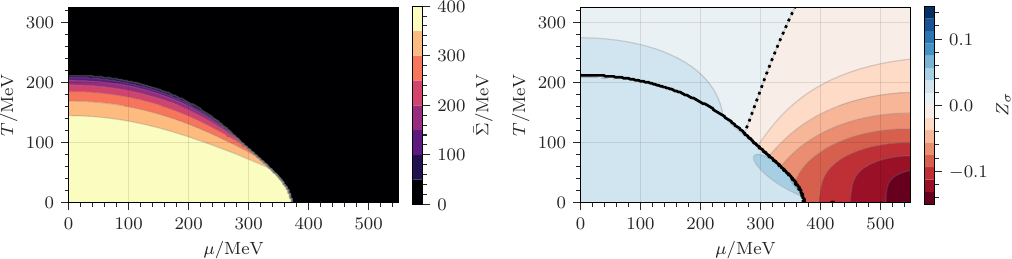}\\[3mm]
			\includegraphics[width=\linewidth]{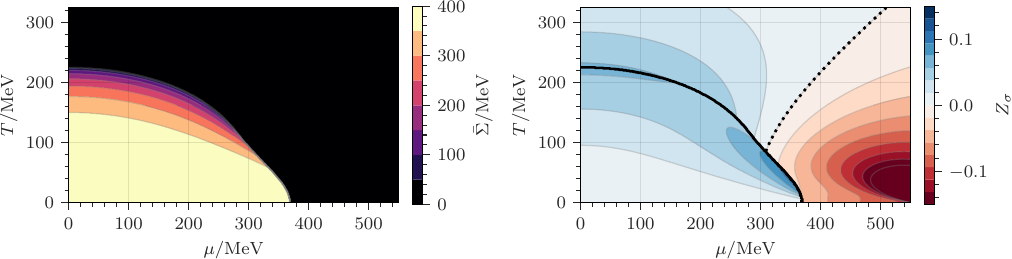}
			\caption{The homogeneous minimum of the effective action \textbf{(left)} and the scalar-wave function renormalization \textbf{(right)} in the \gls{pv} \textbf{(top)} and \gls{sc} \textbf{(bottom)} scheme for $\Mzero = 400\, \text{MeV}$. In the right column, the dotted lines correspond to $\zPhys[\sigma] = 0$ and the solid lines indicate the location of the homogeneous phase transition.}
			\label{fig:mzscannewvacm0400}
		\end{figure*}

		The moat regime appears to be a fairly stable feature of the phase diagram at zero temperature.
		In the following we explore, whether this is also the case for finite temperature and map out the moat regime in the $(T,\mu)$-plane for the \gls{pv} and the \gls{sc} scheme.
		We present results for $\Mzero = 400\, \mathrm{MeV}$, but the relative differences within the schemes, remain mild even for significant variations of $\Mzero$.
		In \cref{fig:mzscannewvacm0400}, we plot the value of $\zPhys[\sigma]$ on the right side alongside the value of the homogeneous minimum $\Mass = \HomMin$ in a color map on the left side for both schemes.
		Again, we obtain fair agreement for the moat regime within the two schemes.
		Starting from the $\zPhys[\sigma] = 0$ line (depicted by the dotted line), the wave function renormalization starts to decrease when increasing $\mu$ at fixed temperature.
		In contrast, when one starts at a fixed $\left(\mu, T\right)$ point within the moat regime, $\zPhys[\sigma]$ grows with increasing $T$ crossing the $\zPhys[\sigma] = 0$ line at a certain point.
		This qualitative behavior is identical within both \glspl{rs} and reminds of the moat regime that is found in the $(1+1)$-dimensional \gls{gn} model \cite{Koenigstein:2021llr}.
		In the \gls{sc} scheme, the moat regime starts at a lower temperature along the homogeneous phase boundary of the \gls{hbp}.
		This is caused by the splitting of the \gls{cep} and the \gls{lp} in the \gls{sc} scheme, as also obtained in the $(\mu, \Mzero)$ phase diagram at zero temperature and discussed in \cref{sec:general_phase_structure}. 
		In the \gls{pv} scheme, one finds $\zPhys[\sigma] \leq 0$ adjacent to the \gls{cep} in the phase diagram, as the \gls{cep} coincides with the \gls{lp} in this scheme \cite{Nickel:2009wj}.
	
\begin{figure}
			\centering
			\includegraphics[width=\linewidth]{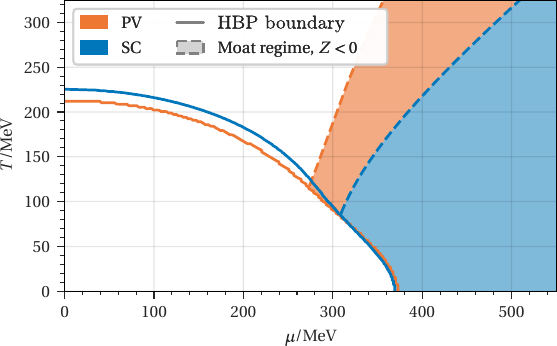}
			\caption{Homogeneous phase boundaries and the moat regime for $\Mzero = 400\, \text{MeV}$ for both the \gls{pv} and the \gls{sc} schemes.}
			\label{fig:mzscancorrectdifferenceregionsnewvacm0400}
		\end{figure}

	   	In \cref{fig:mzscancorrectdifferenceregionsnewvacm0400}, the respective moat regimes of the \gls{pv} and \gls{sc} scheme are compared in a single plot of the $(T,\mu)$-plane.
	   	This demonstrates again the similarity of the moat regime within the two schemes.
	   	Small differences as shown in the plot are expected, since the \gls{njl} model is an effective theory with a dependence on the regularization.
	   	This dependence should, however, be mild to consider the results a reasonably accurate phenomenological description of certain aspects of the related fundamental theory, which is \gls{qcd}.
	   	As the quantitative differences concerning the moat regime within the explored schemes are indeed small, the presented results might be a hint that a moat regime of similar extent also exists in \gls{qcd}. 
	   	In other words, the consistent results in both the $(\mu, \Mzero)$  and the $(T,\mu)$-plane with respect to the regularization suggest that the moat regime is a consequence of the action of the \gls{njl} model and not an artifact associated with the \gls{rs}.

	   	From \cref{eq:zsmall}, the definition of $\z[\phi]$, and the integrals appearing in the two-point function $\gtwo$, one can understand, why the moat regime is more robust than preditions of an \gls{ip}.
	 	$\z[\phi]$ is a quantity exclusively evaluated at $q= 0$. 
	 	Thus, one expects a smaller impact from the \gls{uv} regularization.
	   	In this sense, the evaluation of $\z[\phi]$ is similar to the homogeneous phase boundary, which is also located at values of the chemical potentials of the order of the regulator, but not larger, and which also does not exhibit a sizable \gls{rs} dependence.

	\subsection{The wave-function renormalization at the symmetric point}
	
\begin{figure}
			\centering
			\includegraphics{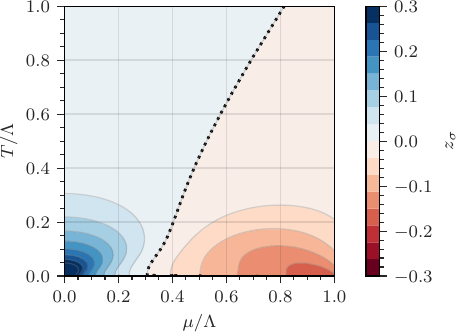}\\[3mm]
				\includegraphics{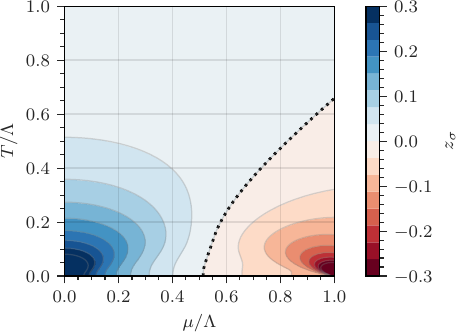}
			\caption{
				The wave-function renormalization $\z[\sigma]$ at the symmetric point $\mstar = 0$ in the $(T,\mu)$-plane for the \gls{pv} \textbf{(top)} and the \gls{sc} \textbf{(bottom)} regularization.
				The dotted line corresponds to the line where $\z[\sigma] = 0$.
			}
			\label{fig:zscansymmetric}
		\end{figure}

		The robustness of the moat regime also stems from the fact that $\z[\phi]$ is independent of the \gls{ff} coupling $\Gcoupling$, but only depends on the dimensionless ratios $\mstar/\regulator$, $T/\regulator$ and $\mu/\regulator$, as can be seen from appropriately rescaling \cref{eq:app:zPVGeneral} and \cref{eq:app:z3DCutoffGeneral} with $\regulator$.
		$\zPhys[\sigma]$ has to be evaluated at $\mstar=\mstarMin$, which is given by the minimization of the effective potential and thus introduces an implicit dependence on $\Gcoupling$, when it is evaluated in the \gls{hbp}.
		In the \gls{sp} and \gls{ip} (where the moat regime is found, compare to \cref{fig:moattzero,fig:mzscannewvacm0400,fig:mzscancorrectdifferenceregionsnewvacm0400}) $\mstar=0$ and, consequently, $\z[\sigma](\mstar=0) = \zPhys[\sigma]$ only depends on $T/\regulator$ and $\mu/\regulator$.
		This is also the reason that the boundaries of the moat regimes in \cref{fig:moattzero} in the lower plot are independent of $\Mzero$, while the boundaries of the \glspl{ip} in \cref{fig:combi}  in the lower plot are a function of $\Mzero$.
		We depict $\z[\sigma](\mstar=0)$ in \cref{fig:zscansymmetric} for the \gls{pv} and the \gls{sc} scheme in the $(\mu/\regulator,T/\regulator)$-plane. 
		Again, we find a similar behavior of $\z[\sigma]$ within both \glspl{rs} in agreement with the previous discussion.
		
		Of course, the region, where $\mstarMin=0$ still depends on the respective value of the regulator $\Lambda$ and the coupling $\Gcoupling$ and, thus, on the selected parameters $\piondecay$ and $\Mzero$.
		Within this region corresponding to the \gls{sp} and \gls{ip}, however, the behavior of $\zPhys[\sigma]$ is universal when rescaling the temperature and chemical potential by the respectively used regulator $\Lambda$.  
		This is also an explanation for the behavior discussed in \cref{sec:moat_zerot}, where the moat regime disconnects from the phase boundary to the \gls{hbp} at small values of $\Mzero$ (see \cref{fig:moattzero}).
		Such small values of $\Mzero$ correspond to large values of $\regulator$.
		This causes a large scaling of \cref{fig:zscansymmetric}, which effectively moves the moat regime to large chemical potentials in units of $\mathrm{MeV}$.
		
	\section{Conclusions}
	\label{sec:conclusions}\glsresetall
	
	    In this work, we studied the \gls{ip}, the moat regime and the corresponding phase diagram in the $3+1$-dimensional \gls{njl} model with a particular focus on their \gls{rs} dependence. 
	    We also explored the dependence on the value of the respective regulator, which implicitly controls the vacuum constituent quark mass $\Mzero$.
	    We fixed the pion decay constant at the physically motivated value $\piondecay=88 \, \mathrm{MeV}$ and compared several common \glspl{rs} at identical values of $\Mzero$ in the range $200 \, \mathrm{MeV}\lesssim \Mzero \lesssim 450\, \mathrm{MeV}$.
	    
	    We found that the extent and shape of the \gls{ip} in the $(\mu,\Mzero)$-plane at $T=0$ drastically differ within each of the considered \glspl{rs}.
	    For example within the \hybridCos{} scheme an \gls{ip} does not even exist. Within the remaining four \glspl{rs}, there is not a single point in the $(\mu,\Mzero)$-plane, where the \glspl{rs} exhibit an \gls{ip} simultaneously. 
	    These discrepancies appear in regions where the chemical potential is of the order of the regulator. 
	    In addition to that the preferred bosonic momenta inside the \glspl{ip} are typically of the order of twice the respective regulator value.
	    This facilitates strong regularization artifacts within all \glspl{rs} as there is no separation of scales of the regulator and physical quantities.
	    Such a separation of scales is, however, important and necessary for a non-renormalizable effective theory like the \gls{njl} model to exhibit a rather weak regulator dependence.
	    We want to explicitly highlight that the separation of scales is already violated by the large values of $\mu$ that are interesting for the investigation of \glspl{ip} and are not necessarily related to the study of inhomogeneous condensates. 
	    In principle, the \gls{rs} dependence could possibly also be present for other (homogeneous) quantities such as, e.g., a baryon density. 
	    However, the effects on the \glspl{ip} are particularily severe as another energy scale, the momentum of the chiral condensate, enters the calculation and is also of the order of or larger than the respective regulator value.
	    This explains why one observes a particularly strong dependence of the \gls{ip} on the \gls{rs} and the respective value of the regulator compared to, e.g., the homogeneous phase boundary.
	    Our results confirm that predictions are not stable and, consequently, not thrustworthy, not even on a qualitative level, in particular considering only results from a single \gls{rs}.
	    For large regulator values, where the required separation of scales is present, an \gls{ip} does not exist independent of the choice of the \gls{rs}.
	    The corresponding $\Mzero$ values are, however, too small to even crudely approximate \gls{qcd}.
	 	   
	    Furthermore, we explored the effect of \gls{nmr} on the shape and extent of the \gls{ip}.
	    We found for the \gls{sc} scheme with \gls{nmr} (where \gls{nmr} has a drastic effect) and the \gls{pv} scheme (where \gls{nmr} has no effect) a qualitative agreement in the shape of the \gls{ip} in the $(\mu,\Mzero)$-plane at $T=0$.
	    While this might be a first step towards a regulator independent result in this model, there is still no obvious physical explanation why the application of \gls{nmr} should yield more thrustworthy predictions for \gls{qcd} than other regularization methods.\footnote{In the \gls{njl} model specifying the \gls{rs} is part of the definition of the model.  
	    Thus, one needs robust evidence that a certain \gls{rs} renders the \gls{njl} model a better effective theory for \gls{qcd} in order to promote results obtained with one \gls{rs} over the ones with other \glspl{rs}. }
	    
	    Moreover, we investigated the moat regime within the \gls{sc} and \gls{pv} schemes in the $(\mu,T)$-plane for various values of $\Mzero$.
	    In constrast to the \gls{ip}, there is fair agreement regarding the extent of the moat regime within the two schemes and it covers a large portion of the \gls{sp}.
	    This indicates that the existence of the moat regime could be a rather robust feature of the phase diagram of the \gls{njl} model, largely independent of the used \gls{rs}.
	    Even though the chemical potential is again of the order of the respective regulator, the moat regime has a much weaker dependence than \glspl{ip} as it is determined by the wave function renormalization, which is calculated at vanishing external bosonic momenta compared to the \gls{ip}, where the relevant momenta are of the order of $2 \regulator$.
	    Compared to the \gls{ip} there is no further energy scale entering the computation of the wave function renormalization and, thus, the violation of the separation of scales is not further enhanced.\footnote{The quantity violating the separation of scales is the chemical potential. This is also the case for, e.g., the homogeneous phase boundary which is also rather stable with respect to variations of the \gls{rs} compared to the \gls{ip}.}
	    This is a likely explanation why the moat regime is rather stable with respect to variations of the \gls{rs} compared to the \gls{ip}.
	
		To summarize, we have presented numerical results and theoretical arguments	indicating that results on \glspl{ip} in the $3+1$-dimensional \gls{njl} model are strongly \gls{rs} dependent and, thus, do not provide evidence for the existence or absence of an \gls{ip} in \gls{qcd}.
		On the other hand, the stability of the moat regime in the $3+1$-dimensional \gls{njl} model with respect to variations of the \gls{rs} and the value of the regulator could indicate its existence in \gls{qcd}.
	    First evidence from \Rcite{Fu:2019hdw} applying the functional renormalization group to \gls{qcd}, which found a moat regime at finite $\mu$ and $T$, supports this conclusion.

    \subsection{Outlook}
	    The observed \gls{rs} dependence of the \gls{ip} could also play an critical role in \gls{njl} model investigations of \glspl{ip} with non-vanishing magnetic fields \cite{Ferrer:2019zfp,Gyory:2022hnv,Ferrer:2023xvl}, under rotation \cite{TabatabaeeMehr:2023tpt,Chen:2024tkr} or in nuclear matter models \cite{Takahashi:2001jq, Pitsinigkos:2023xee}.
		Also in these scenarios, it is important to carefully consider and study different \glspl{rs} and their influence on the phase diagram. 
		In order to make robust statements regarding the \gls{ip} using \gls{njl}-type models, one would need to figure out which \gls{rs} renders the \gls{njl} model the best description of finite-density \gls{qcd}. 
		This, however, should be guided by finite density \gls{qcd} phenomenology, for which unfortunately little evidence exists.
	
	    To solidify the finding that the moat regime is a robust feature of the \gls{njl} model, a first step would be to study the moat regime with further \glspl{rs}.
	    Also, it would be interesting to investigate the existence of the moat regime in other (possible more elaborate) model approaches, such as, e.g., the Polyakov-loop quark-meson model, in order to investigate the robustness of the moat regime within different models. 
	    If there is universal behavior of strongly-interacting fermions with respect to the moat regime, one could make use of the mesonic dispersion relations found in the moat regimes for identifying experimental signals such as in \Rcite{Pisarski:2021qof,Rennecke:2023xhc, Fukushima:2023tpv}.
    
    \section*{Data availability}
    
    The plotted data is available via \Rcite{pannulloDataCodeInhomogeneous2024}.

	\begin{acknowledgments}
		We thank M.~Buballa, H.~Gholami, A.~Königstein, J.~Lenz, M.~Mandl, G.~Markó, Z.~Nussinov, M.~Ogilvie, R.~Pisarski, F.~Rennecke, S.~Schindler, A.~Sciarra and A.~Wipf for fruitful discussions related to this work.
		L. P. thanks G. Endr\H odi for valuable discussions and for his general support at the faculty of physics at the University of Bielefeld.
		We acknowledge the support of the \textit{Deutsche Forschungsgemeinschaft} (DFG, German Research Foundation) through the collaborative research center trans-regio  CRC-TR 211 ``Strong-interaction matter under extreme conditions''-- project number 315477589 -- TRR 211.	
		M.~Wa. ~acknowledges support by the Heisenberg Programme of the  \textit{Deutsche Forschungsgemeinschaft} (DFG, German Research Foundation) under grant number 399217702.
		L.P.~and M.Wi.~acknowledge the support of the \textit{Helmholtz Graduate School for Hadron and Ion Research} and of the \textit{Giersch Foundation}.
		
	\end{acknowledgments}
	\bibliography{main}
	\appendix
	\section{Parameter fitting\label{app:formulas_parameterfitting}}
		In this appendix, we discuss how the parameters $G$ and $\regulator$ in the \gls{njl} model (see the discussion in the last paragraph of \cref{sec:hompot} and in \cref{sec:parameterfitting}) are fitted to the constituent quark mass $\Mzero$ and the pion decay constant $\piondecay$ in the vacuum.
		We use the gap equation \eqref{eq:gap} to relate $G$ to the ratio $\Mzero / \regulator$ and the vacuum-to-one-pion axial vector matrix element to express $\piondecay / \Mzero$ as a function of $\Mzero / \regulator$, as proposed in \Rcite{Klevansky:1992qe}.
		These two relations itself and, consequently, the solution for $G$ and $\regulator$ depends on the respective \gls{rs}.
		In the following, we present the relevant expressions for this computation for each of the five used \glspl{rs}.
		
	\subsection{Pauli-Villars}
		Computing the vacuum-to-one-pion axial vector matrix element yields
		\begin{align}
			\frac{\piondecay^2}{\Mzero^2} = - \frac{\Nbar}{8 \uppi^2} \pvsumck[0] \ln \left(\frac{\Mass_{0,k}^2}{\regulatorPV^2}\right), \label{eq:Appendix:fpiPV}
		\end{align}
		while the gap equation \eqref{eq:gap} gives 
		\begin{align}
			\frac{1}{\Gcoupling} = \frac{\Nbar}{8 \uppi^2} \pvsumck[0] \Mass_{0,k}^2 \ln \left(\frac{\Mass_{0,k}^2}{\regulatorPV^2}\right)  , \label{eq:Appendix:CouplingPV}
		\end{align}
		where $\Mass_{0,k}^2=\Mzero^2+\alpha_k\regulatorPV^2$. 
		See below \cref{eq:fermi_action} for the definition of $\Nbar$.
	
	\subsection{Spatial Momentum Cutoff}
		For the spatial momentum cutoff, we find
		\begin{align}
			\frac{\piondecay^2}{\Mzero^2} = \frac{\Nbar}{16 \uppi^2} \left[\arsinh\left(\frac{\regulatorSC}{\Abs{\Mzero}}\right) - \left(\frac{\Mzero^2}{\regulatorSC^2}+1 \right)^{-\frac{1}{2}}\right], \label{eq:Appendix:fpiSC}
		\end{align}
		and
		\begin{align}
			\frac{1}{\Gcoupling} = \frac{\Nbar}{4 \uppi^2} \left[\regulatorSC\sqrt{\Mzero^2 + \regulatorSC^2}  - \arsinh\left(
			\frac{\regulatorSC}{\Abs{\Mzero}}\right) \Mzero^2\right], \label{eq:Appendix:CouplingSC}
		\end{align}
		respectively.
		
	\subsection{Lattice Regularization}
	\label{sec:Appendix:ParameterFittingLattice}
		The formulas for the lattice regularizations differ exclusively in their expression for the energy $E$ (see \cref{eq:latt_energy}). 
		Therefore, we do not need to distinguish between the SLAC and Hybrid discretization here.
		The quantities of the parameter fitting are given by
		\begin{align}
			\frac{\piondecay^2}{\Mzero^2} = \frac{\Nbar}{  4 \uppi^4}&  
			\int_{0}^{\regulatorLFT} \!\!\!\!\dr[p_1] 
			\int_{0}^{\regulatorLFT} \!\!\!\!\dr[p_2] 
			\int_{0}^{\regulatorLFT} \!\!\!\!\dr[p_3] \times \nonumber \\ &\times \left[\frac{\arctan\left(\frac{\regulatorLFT}{E}\right)}{ E^3}+\frac{ \regulatorLFT}{ E^2  (E^2 + \regulatorLFT^2)}\right]
			\label{eq:Appendix:fpiLattice}
		\end{align}
		and
		\begin{align}
			\frac{1}{\Gcoupling} =  \frac{2\Nbar}{  \uppi^4} 
			\int_{0}^{\regulatorLFT} \!\!\!\!\dr[p_1] 
			\int_{0}^{\regulatorLFT} \!\!\!\!\dr[p_2] 
			\int_{0}^{\regulatorLFT}  \!\!\!\!\dr[p_3] \frac{1}{E}   \arctan\left(\frac{E}{\regulatorLFT}\right) 
			. \label{eq:Appendix:CouplingLattice}
		\end{align}
		The energies for the respective lattice regularization are
		\begin{equation}
			E = \sqrt{\sum_{i=1}^3 \fermionDisp_X^2(p_i) + \mstar^2 } \label{eq:latt_energy}
		\end{equation}
		where $\fermionDispSLAC$ is defined in \cref{eq:5:SLACDisp} and 
		\begin{equation}
			\fermionDisp_{\text{Hybrid}}(p_\mu) = \delta_{\mu,0}\fermionDispSLAC(p0) + \delta_{\mu,j} \sin(\mathbf{p}_j a). \label{eq:HybridDisp}
		\end{equation} 
		
	\widetext{
	\section{Homogeneous effective potential\label{app:formulas_hompout}}
	
		Here, we give the relevant expressions for the evaluation of the homogeneous effective potential \eqref{eq:hompotellZero} by specifying $\ellZero$ for each of the five \glspl{rs} used.
		The unregularized and therefore divergent expression is given by 
		\begin{equation}
			\ellZero\left(\mstar,T,\mu\right)={} \frac{1}{2\uppi^2} \int_{0}^{\infty} \dr[p] p^2 \left[E + \tfrac{1}{\beta} \ln \left(1+\eu^{-\beta(E+\mu)}\right)+\tfrac{1}{\beta} \ln \left(1+\eu^{-\beta(E-\mu)}\right)\right] \label{eq:ellZeroDef}
		\end{equation}
		with $E = \sqrt{p^2 + \mstar^2}$.
		\subsection{Pauli-Villars}
		For arbitrary $\mstar, T, \mu$, one finds 
	
		\begin{align}
			\ellZero\left(\mstar,T,\mu\right)={}& \frac{1}{2\uppi^2} \int_{0}^{\infty} \dr[p] p^2 \pvsumck[0] \left[E_k + \tfrac{1}{\beta} \ln \left(1+\eu^{-\beta(E_k+\mu)}\right)+\tfrac{1}{\beta} \ln \left(1+\eu^{-\beta(E_k-\mu)}\right)\right] = \nonumber\\
			={}&
			\frac{1}{32\uppi^2} \Bigg\{  \pvsumck[0]  \Mass_k^4 \ln\left(\frac{\Mass_k^2}{\regulatorPV^2}\right)  + 16 \int_{0}^{\infty} \dr[p] \frac{p^4}{3} \pvsumck[0]  \frac{\distributionF{E_k}+\distributionFBar{E_k}}{E_k}\Bigg\} ,
		\end{align}
		where we used partial integration and defined the Fermi-Dirac distribution functions 
		\begin{equation}
			\distributionF{x} = \frac{1}{1 + \e^{\beta (x-\mu)}}, \; \distributionFBar{x} = \frac{1}{1 + \e^{\beta (x+\mu)}}. \label{eq:Fermi_distributions}
		\end{equation}
	
		For $\mu = T = 0$, we find
		\begin{align}
			\ellZero\left(\mstar,T=0,\mu=0\right)
			={}&
			\frac{1}{32\uppi^2}  \pvsumck[0]  \Mass_k^4 \ln\left(\frac{\Mass_k^2}{\regulatorPV^2}\right)
		\end{align}
		implying 
		\begin{align}
			\ellZero\left(\mstar=0,T=0,\mu=0\right)
			={}&
			\frac{\regulatorPV^4 }{32\uppi^2}  \pvsumck[1]  \alpha_k^2 \ln\left(\alpha_k\right)  .
		\end{align}
		using that $\Mass_k = \alpha_k$ for $\mstar = 0$.
		Moreover, we find
		\begin{align}
			\ellZero\left(\mstar,T=0,\mu\right)
			={}&
			\frac{1}{32\uppi^2}   \pvsumck[0] \Bigg\{ \Mass_k^4 \ln\left(\frac{\Mass_k^2}{\regulatorPV^2}\right)  + \Theta\left(\mub_k^2\right)\frac{2 }{3} \left[3 \Mass_k^2 \arsinh\left(\frac{\mub_k}{\abs{\Mass_k}}\right) + \mub_k \left(2 \Abs{\mu}^3 - 5 \Mass_k^2 \Abs{\mu}\right)  \right] \Bigg\} ,
		\end{align}
		and
		\begin{align}
			\begin{split}
				\ellZero\left(\mstar=0,T=0,\mu\right)={}&\frac{1}{32\uppi^2}    \Bigg\{ \regulatorPV ^4\pvsumck[1]  \alpha_k^2 \ln\left(\alpha_k\right) +\Theta\left(\mu^2\right) \frac{4\mu^4 }{3} +\\
				&\hphantom{\frac{1}{32\uppi^2}    \Bigg\{}+ \pvsumck[1] \Theta\left(\mub_k^2\right) \frac{2 }{3} \left[3 \Mass_k^2 \arsinh\left(\frac{\mub_k}{\abs{\Mass_k}}\right) + \mub_k \left(2 \Abs{\mu}^3 - 5 \Mass_k^2 \mu\right)  \right] \Bigg\} ,
			\end{split}
		\end{align}
		where 
		\begin{align}
			\mub_k = \sqrt{\mu^2 - \Mass_k^2}.
		\end{align}
}
	\subsection{Spatial Momentum Cutoff}
		For the \gls{sc} scheme, one obtains
		\begin{align}
			\ellZero\left(\mstar,T,\mu\right)={}& \frac{1}{2\uppi^2} \int_{0}^{\regulatorSC} \dr[p] p^2 \left[E + \tfrac{1}{\beta} \ln \left(1+\eu^{-\beta(E+\mu)}\right)+\tfrac{1}{\beta} \ln \left(1+\eu^{-\beta(E-\mu)}\right)\right] = \nonumber\\
			\begin{split}
				={}&
				\frac{1}{16\uppi^2} \Bigg\{ \regulatorSC\sqrt{\regulatorSC^2+\mstar^2}\left(2\regulatorSC^2+\mstar^2\right)-\mstar^4 \arsinh\left(\frac{\regulatorSC}{\abs{\mstar}}\right)  +\\
				& \hphantom{\frac{1}{16\uppi^2} \Bigg\{ }+ 8 \int_{0}^{\regulatorSC} \dr[p] p^2 \left[ \tfrac{1}{\beta} \ln \left(1+\eu^{-\beta(E+\mu)}\right)+\tfrac{1}{\beta} \ln \left(1+\eu^{-\beta(E-\mu)}\right)\right]\Bigg\} .
			\end{split}
		\end{align}
		
		For $\mu = T = 0$, one finds the closed form expression
		\begin{align}
			\ellZero\left(\mstar,T=0,\mu=0\right)={}& 	\frac{1}{16\uppi^2}\left[\regulatorSC\sqrt{\regulatorSC^2+\mstar^2}\left(2\regulatorSC^2+\mstar^2\right)-\mstar^4 \arsinh\left(\frac{\regulatorSC}{\abs{\mstar}}\right) \right]
		\end{align}
		which reduces to 
		\begin{align}
			\ellZero\left(\mstar=0,T=0,\mu=0\right)={}& 	\frac{\regulatorSC^4}{8\uppi^2}
		\end{align}
		for $\mstar = 0$.
		
		Moreover, one finds
		\begin{align}
			&\ellZero\left(\mstar,T=0,\mu\right)=\nonumber\\
			={}& \ellZero\left(\mstar,T=0,\mu=0\right) -  \frac{\Theta \left(\mub^2\right)}{16\uppi^2}  \int_{0}^{P=\min\left(\regulatorSC,\mub\right)} \hspace{-3ex} \dr[p] p^2 \left(E-\abs{\mu}\right) =  \nonumber\\
			={}&\ellZero\left(\mstar,T=0,\mu=0\right) -  \frac{\Theta \left(\mub^2\right)}{16\uppi^2} \left[P\sqrt{P^2+\mstar^2}\left(2P^2+\mstar^2\right)-\mstar^4 \arsinh\left(\frac{P}{\abs{\mstar}}\right) - \frac{\Abs{\mu} P^3}{3} \right] = \nonumber\\
			={}&
			\frac{1}{16\uppi^2} \begin{cases}
				\regulatorSC\sqrt{\regulatorSC^2+\mstar^2}\left(2\regulatorSC^2+\mstar^2\right)-\mstar^4 \arsinh\left(\frac{\regulatorSC}{\abs{\mstar}}\right) & \ifc \mub=0,\\
				\regulatorSC\sqrt{\regulatorSC^2+\mstar^2}\left(2\regulatorSC^2+\mstar^2\right)-\mstar^4 \arsinh\left(\frac{\regulatorSC}{\abs{\mstar}}\right)\\
				-\mub\abs{\mu}^3+\mstar^4 \arsinh\left(\frac{\mub}{\abs{\mstar}}\right) + \frac{2}{3} \Abs{\mu} \mub^3& \ifc 0<\mub^2<\regulatorSC^2 ,\\
				\frac{\Abs{\mu} \regulatorSC^3}{3}& \ifc \mub^2>\regulatorSC^2
			\end{cases},
		\end{align}
				where 
		\begin{align}
			\mub = \sqrt{\mu^2 - \mstar^2},
		\end{align}
		and
		\begin{align}
			\ellZero\left(\mstar=0,T=0,\mu\right)={}& \frac{\regulatorSC^4}{8\uppi^2} -  \frac{\Theta \left(\mu^2\right)}{16\uppi^2}  \int_{0}^{P=\min\left(\regulatorSC,\abs{\mu}\right)} \hspace{-3ex} \dr[p] p^2 \left(p-\abs{\mu}\right) =  \nonumber\\
			={}&\frac{1}{8\uppi^2} \left\{\regulatorSC^4- \Theta \left(\mu^2\right) \left[P^4 - \frac{4}{3}\Abs{\mu} P^3 \right]\right\} =\nonumber\\
			={}&\frac{1}{8\uppi^2}  \begin{cases}
				\regulatorSC^4 & \ifc \mu=0,\\
				\regulatorSC^4 +\frac{1}{3} \mu^4&  0<\mu^2\leq\regulatorSC^2 ,\\
				\frac{4}{3} \Abs{\mu} \regulatorSC^3 & \mu^2>\regulatorSC^2
			\end{cases}.
		\end{align}

	\subsection{Lattice Regularization}
	\label{sec:Appendix:njlueffLattice}
		The formulas for the lattice regularizations differ exclusively in their expression for the energy $E$ (see \cref{eq:latt_energy}). 
		Therefore, we do not need to distinguish between the SLAC and Hybrid discretization here.
		The resulting expression in the zero temperature limit is 
		\begin{align}
			\ellZero\left(\mstar,T=0,\mu\right)={}& \frac{8}{(2 \uppi)^4}   
			\int_{0}^{\regulatorLFT} \!\!\!\! \dr[p_1] \!\!
			\int_{0}^{\regulatorLFT} \!\!\!\! \dr[p_2] \!\! 
			\int_{0}^{\regulatorLFT} \!\!\!\! \dr[p_3] \Big[-4\regulatorLFT- 2 \Abs{\mu} \arctantwo\left(2 \Abs{\mu} \regulatorLFT , E^2 - \mu^2 + \regulatorLFT^2\right)+\nonumber\\
			& 
			+ 2 E \arctantwo\left(2  E \regulatorLFT, E^2 - \mu^2 - \regulatorLFT^2\right)
			+ \regulatorLFT  \ln\left(4 \mu^2\regulatorLFT^2 + \left( E^2 - \mu^2 + \regulatorLFT^2\right)^2\right)+\nonumber\\
			&- \Theta(\mu^2-E^2) 2  \uppi  (E - \Abs{\mu})\Big],
		\end{align}
		where 
		\begin{align}
			\arctantwo(y,x) = \begin{cases}
				\arctan\left(\frac{y}{x}\right) & \ifc x>0, \\
				\arctan\left(\frac{y}{x}\right) + \uppi & \ifc x<0 \text{ and } y\geq 0, \\
				\arctan\left(\frac{y}{x}\right) - \uppi & \ifc x<0 \text{ and } y< 0, \\
				+\frac{\uppi}{2} & \ifc x=0 \text{ and } y> 0, \\		
				-\frac{\uppi}{2} & \ifc x=0 \text{ and } y< 0, \\
				\text{undefined} & \ifc x=0 \text{ and } y= 0.	
			\end{cases}.
		\end{align}
	
	\section{Stability analysis\label{app:formulas_stability}}
		For the \gls{sc} and \gls{pv} scheme, the bosonic two-point function is split up into the $q$-independent part $\ellOne$ and a $q$-dependent part $\EllTwo$. 
		For the two-point function of bosonic field $\phi_i=(\sigma,\pmb{\pi})_i$, one finds
		\begin{equation}
			\Gamma^{(2)}_{\phi_i}(\mathbf{q}, \mstar, \mu, T) = \frac{1}{2 G} - \Nbar \ellOne(\mstar, \mu, T) + \Nbar \EllTwo[,\phi_i](\mathbf{q}, \mstar, \mu, T) \label{eq:app:gamma2splitup}
		\end{equation}
		with 
		\begin{equation}
			\ellOne = \int \tfrac{\dr^3 p}{(2 \uppi)^3} \tfrac{1}{\beta}  \sum_{n = -\infty}^{\infty} \frac{1}{\left(\nu_n - \i \mu\right)^2 + E^2} = \int \tfrac{\dr^3 p}{(2 \uppi)^3} \frac{1 - \distributionF{E} - \distributionFBar{E}}{E} \label{eq:ellOne}
		\end{equation}
		and 
		\begin{align}
			 \EllTwo[,\phi_i](\mathbf{q}, \mstar, \mu, T) &= \int \tfrac{\dr^3 p}{(2 \uppi)^3} \tfrac{1}{\beta} \sum_{n = -\infty}^{\infty} \frac{\mathbf{p} \cdot \mathbf{q} + \mathbf{q}^2 + 2 \delta_{i,0} \mstar^2}{\left[\left(\nu_n - \i \mu\right)^2 + E_{\mathbf{p}}^2\right] \left[\left(\nu_n - \i \mu\right)^2 + E_{\mathbf{p}+ \mathbf{q}}^2\right] } = \\ \nonumber
			 &= \int \tfrac{\dr^3 p}{(2 \uppi)^3} \frac{\mathbf{p} \cdot \mathbf{q} + \mathbf{q}^2 + 2 \delta_{i,0} \mstar^2}{\mathbf{q}^2 + 2 \mathbf{p} \cdot \mathbf{q}} \times \left[\frac{1 - \distributionF{E} - \distributionFBar{E}}{E} - \frac{1 - \distributionF{E_{\mathbf{p}+ \mathbf{q}}} - \distributionFBar{E_{\mathbf{p}+ \mathbf{q}}}}{E_{\mathbf{p}+ \mathbf{q}}}\right].
			 \label{eq:Ell2}
		\end{align}
	\subsection{Pauli-Villars}
	\label{sec:Appendix:PVstab}
		\subsubsection{The momentum independent integral $\ellOne$}
		In the \gls{pv} scheme, we have
		\begin{align}
			\ellOneArgs{\mstar}{\mu}{T}  = \frac{1}{4 \uppi^2} \pvsumck[0]\left\{\Mass_k^2 \ln\left(\frac{\Mass_k^2}{\regulatorPV^2}\right) - \int_0^\infty  \dr[p] p^2 \frac{\distributionF{E_k}+\distributionFBar{E_k}}{2 E_k} \right\}.
		\end{align}
		The vacuum contribution reads
		\begin{align}
			\ellOneArgs{\mstar}{\mu=0}{T=0}  = \frac{1}{4 \uppi^2} \pvsumck[0]\Mass_k^2 \ln\left(\frac{\Mass_k^2}{\regulatorPV^2}\right),
		\end{align}
		where taking $\mstar \rightarrow 0$ yields
		\begin{align}
			\ellOneArgs{\mstar=0}{\mu=0}{T=0}  = \frac{\regulatorPV^2}{4 \uppi^2} \pvsumck[1] \alpha_k  \ln\left(\alpha_k\right).
		\end{align}
		 
		At zero temperature, we find 
		\begin{align}
			\ellOneArgs{\mstar}{\mu}{T=0}  = \frac{1}{4 \uppi^2} \pvsumck[0]\left\{\Mass_k^2 \ln\left(\frac{\Mass_k^2}{\regulatorPV^2}\right) - \frac{\Theta\left(\mub_k^2\right)}{2} \left[\Abs{\mu} \mub_k - \Mass_k^2 \arsinh\left(\frac{\mub_k}{\abs{\Mass_k}}\right)\right] \right\}
		\end{align}
		and for $\mstar=0$ 
		\begin{align}
			\ellOneArgs{\mstar=0}{\mu}{T=0}  = \frac{1}{4 \uppi^2} \left\{ \pvsumck[1 ]\alpha_k  \ln\left(\alpha_k\right)  - \Theta\left(\mu^2\right)\frac{\mu^2}{2} - \pvsumck[1]\frac{\Theta\left(\mub_k^2\right)}{2} \left[\Abs{\mu} \mub_k - \Mass_k^2 \arsinh\left(\frac{\mub_k}{\abs{\Mass_k}}\right)\right] \right\}.
		\end{align}
	
	\subsubsection{The Momentum Dependent Integral $\EllTwo$}
		We perform a shift in the momentum integral of $\EllTwo[,\phi_i]$ and obtain
		\begin{align}
			\EllTwo[,\phi_i]\left(\x{q},\mstar,\mu,T\right)
			={}& \left(\x{q}^2 + \kron{i}{1} 4 \mstar^2\right)  \int  \intMeasureOverPiNew[3]{p} \frac{1 }{2 \vecb{p} \cdot \vecb{q} - q^2} \frac{\OneMinusnsDef{E} }{ 2E } \equiv \left(\x{q}^2 + \kron{i}{1} 4 \mstar^2\right) \ellTwo\left(\x{q},\mstar,\mu,T\right), \label{eq:appendix:elltwodefinition}
		\end{align}
		where only $\ellTwo$ will be regulated using the \gls{pv} scheme to be consistent with existing literature \cite{Carignano:2014jla, Buballa:2020xaa}.
		We, thus, obtain
		\begin{align}
			\EllTwo[,\phi_i]\left(q,\mstar,\mu,T\right)
			={}&  - \frac{ \left(q^2 + \kron{i}{0} 4 \mstar^2\right)}{ (4 \pi )^2} \sum_{k=0}^{\NPV} c_k \, \Bigg\{ \Bigg[ \frac{1}{2}   \ln\left(\frac{\Mass_k^2}{\regulatorPV^2}\right)  +  \sqrt{1+\frac{4\Mass_k^2}{q^2}} \, \arcoth \left(\sqrt{1+\frac{4\Mass_k^2}{q^2}}\right)	 \Bigg] +  \nonumber\\
			&\hphantom{- \frac{ \left(\x{q}^2 + \kron{i}{0} 4 \mstar^2\right)}{ (4 \pi )^2} \sum_{k=0}^{\NPV} c_k \, \Bigg\{}
			-   \int_{0}^{\infty} \dr p \,  \frac{p}{q}\, \frac{\distributionF{E_k}+\distributionFBar{E_k}}{ \E{k}}  \ln \left|\frac{2p-q}{2p+q}\right|	 \Bigg\}  \label{eq:Appendix:NJLEll2PVGeneral}
		\end{align}
		and taking the zero temperature limit yields
		\begin{align}
			&\EllTwo[,\phi_i]\left(q,\mstar,\mu,T=0\right) =\nonumber\\
			\begin{split}
				={}&  - \frac{ \left(q^2 + \kron{i}{0} 4 \mstar^2\right)}{ (4 \pi )^2} \sum_{k=0}^{\NPV} c_k \, \Bigg\{ \Bigg[ \frac{1}{2}  \ln\left(\frac{\Mass_k^2}{\regulatorPV^2}\right)  +  \sqrt{1+\frac{4\Mass_k^2}{q^2}} \, \arcoth \left(\sqrt{1+\frac{4\Mass_k^2}{q^2}}\right)	 \Bigg] +  \\
				&
				-  \Theta\left(\mub_k^2\right) \Bigg[ - \artanh \left(\frac{\mub_k}{|\mu|}\right) + \frac{1}{2}\sqrt{1+\frac{4\Mass_k^2}{q^2}}\, \ln \left|\tfrac{q |\mu|+ \mub_k \sqrt{q^2+4\Mass_k^2}}{q |\mu|- \mub_k \sqrt{q^2+4\Mass_k^2}}\right| - \frac{\Abs{\mu}}{q}\ln \left|\frac{2\mub_k+q}{2\mub_k-q}\right| \Bigg] \Bigg\} 	\ 	 		 	  	.
			\end{split}\label{eq:Appendix:NJLPVTZero}
		\end{align}
		
		In the \gls{sp}, one finds
		\begin{align}
			&\EllTwo[,\phi_i]\left(q,\mstar=0,\mu,T=0\right) = \nonumber\\
			\begin{split}
				={}&  - \frac{ q^2}{ (4 \pi )^2} \, \Bigg\{ \ln \Abs{\frac{q}{\regulatorPV}} + \pvsum[1]\Bigg[ \frac{1}{2}  \ln\left(\frac{\Mass_k^2}{\regulatorPV^2}\right)  +   \sqrt{1+\frac{4\Mass_k^2}{q^2}}\, \arcoth \left( \sqrt{1+\frac{4\Mass_k^2}{q^2}}\right)	 \Bigg] +  \\
				&-  \frac{1}{2q} \left[(2\Abs{\mu}+q)\ln \Abs{\frac{2\Abs{\mu}+q}{\regulatorPV}}+(2\Abs{\mu}-q)\ln \Abs{\frac{2\Abs{\mu}-q}{\regulatorPV}}\right] +\\
				&
				-  \pvsum[1] \Theta\left(\mub_k^2\right) \Bigg[ - \artanh \left(\frac{\mub_k}{|\mu|}\right) +  \frac{1}{2}\sqrt{1+\frac{4\Mass_k^2}{q^2}}\, \ln \left|\tfrac{|q\mu|+ \mub_k \sqrt{q^2+4\Mass_k^2}}{|q\mu|- \mub_k \sqrt{q^2+4\Mass_k^2}}\right| - \frac{|\mu|}{q} \ln \left|\frac{2\mub_k+q}{2\mub_k-q}\right| \Bigg] \Bigg\} \	.
			\end{split}					 	 		 	  	
			\label{eq:AppendiX:EllTwoPVMZeroTZero}
		\end{align}
		
		In the limit of vanishing external momentum $q$, one obtains
		\begin{align}
			\EllTwo[,\phi_i]\left(q=0,\mstar,\mu,T\right)
			={}&  - \frac{ \kron{i}{0} 4 \mstar^2}{ (4 \pi )^2} \, \pvsumck[0] \, \Bigg\{  \frac{1}{2}   \ln\left(\frac{\Mass_k^2}{\regulatorPV^2}\right)  
			-   \int_{0}^{\infty} \dr[p] \frac{\distributionF{E_k}+\distributionFBar{E_k}}{ \E{k}}   \Bigg\}	
		\end{align}
		and in the zero temperature limit one has
		\begin{align}
			\EllTwo[,\phi_i]\left(q=0,\mstar,\mu,T=0\right)
			={}&  - \frac{ \kron{i}{0} 4 \mstar^2}{ (4 \pi )^2} \, \pvsumck[0] \, \Bigg\{  \frac{1}{2}   \ln\left(\frac{\Mass_k^2}{\regulatorPV^2}\right)  
			-  \Theta\left(\mub_k^2\right) \arsinh\left(\frac{\mub_k}{\abs{\mstar}}\right) \Bigg\}	
		\end{align}
		in consistency with the zero $q$ limit of \cref{eq:Appendix:NJLPVTZero}.
		Taking $q= \mstar = 0$, one obtains from \cref{eq:appendix:elltwodefinition}
		\begin{align}
			\EllTwo[,\phi_i]\left(q=0,\mstar=0,\mu,T\right)=0.
		\end{align}
	\subsection{Spatial Momentum Cutoff}
	\label{sec:Appendix:SCstab}
	\subsubsection{The Momentum Independent Integral $\ellOne$}
		For the \gls{sc} scheme, one finds
		\begin{align}
			\ellOneArgs{\mstar}{\mu}{T}
			& = \frac{1}{8 \uppi^2} \left\{ \regulatorSC \sqrt{\mstar^2+\regulatorSC^2}-\arsinh\left(\tfrac{\regulatorSC}{\abs{\mstar}}\right) \mstar^2 -\int_0^{\regulatorSC}  \dr[p] p^2 \frac{\distributionF{E}+\distributionFBar{E}}{E} \right\},
		\end{align}
		which contains the vacuum contribution
		\begin{align}
			\ellOneArgs{\mstar}{\mu=0}{T=0} & = \frac{1}{8 \uppi^2} \left\{ \regulatorSC \sqrt{\mstar^2+\regulatorSC^2}-\arsinh\left(\tfrac{\regulatorSC}{\abs{\mstar}}\right) \mstar^2   \right\}.
		\end{align}	
		For $\mstar = T = \mu = 0$ one has
		\begin{align}
			\ellOneArgs{\mstar=0}{\mu=0}{T=0} & = \frac{\regulatorSC^2}{8 \uppi^2} 	.	
		\end{align}	
		
		At zero temperature, one can give a closed form expression 
		\begin{align}
			\ellOneArgs{\mstar}{\mu}{T=0}  ={}& \frac{1}{8 \uppi^2} \left\{ \regulatorSC \sqrt{\mstar^2+\regulatorSC^2}-\arsinh\left(\tfrac{\regulatorSC}{\abs{\mstar}}\right) \mstar^2 -\int_0^{\regulatorSC}\dr[p]  \frac{p^2}{E}\ 	\Theta\left(\mu^2-E^2\right)\right\} = \nonumber \\
			\begin{split}
				={}& \tfrac{1}{8 \uppi^2} \Bigg\{ \regulatorSC \sqrt{\mstar^2+\regulatorSC^2}-\arsinh\left(\tfrac{\regulatorSC}{\abs{\mstar}}\right) \mstar^2 +\nonumber\\
				& \hphantom{\tfrac{1}{8 \uppi^2} \Bigg\{}- \Theta\left(\mub^2\right) \left[P 	\sqrt{\mstar^2+P^2}-\arsinh\left(\tfrac{P}{\abs{\mstar}}\right) \mstar^2\right]^{P=\min\left(\regulatorSC,\mub\right)}  \Bigg\} =
			\end{split}\\
			={}&\frac{1}{8 \uppi^2} \begin{cases}
				\regulatorSC \sqrt{\mstar^2+\regulatorSC^2}-\arsinh\left(\tfrac{\regulatorSC}{\abs{\mstar}}\right) \mstar^2 &\ifc \mub^2=0,\\
				\regulatorSC \sqrt{\mstar^2+\regulatorSC^2}-\arsinh\left(\tfrac{\regulatorSC}{\abs{\mstar}}\right) \mstar^2 -\mub\abs{\mu}+\arsinh\left(\tfrac{\mub}{\abs{\mstar}}\right) \mstar^2 &\ifc 0<\mub^2<\regulatorSC^2,\\
				0 &\ifc \mub^2>\regulatorSC^2
			\end{cases}
		\end{align}
		and in the limit of vanishing $\mstar$
		\begin{align}
			\ellOneArgs{\mstar=0}{\mu}{T=0} ={}& \tfrac{1}{8 \uppi^2} \Bigg\{ \regulatorSC^2 - \Theta\left(\mu^2\right) \Big[P^2 	\Big]^{P=\min\left(\regulatorSC,\abs{\mu}\right)}  \Bigg\}=\\
			={}&\frac{1}{8 \uppi^2}\begin{cases}
				\regulatorSC^2  &\ifc \mu^2=0,\\
				\regulatorSC^2  -\mu^2 &\ifc 0<\mu^2<\regulatorSC^2,\\
				0 &\ifc \mu^2>\regulatorSC^2.
			\end{cases}
		\end{align}
	
	\subsubsection{The Momentum Dependent Integral $\ellTwo$}
		The evaluation of $\EllTwo$ differs from the \gls{pv} scheme significantly, because shifts in the momentum integral would shift the integration bound $\regulatorSC$.
		One finds
		\begin{align}
			\EllTwo[,\phi_i]\left(q,\mstar,\mu,T=0\right)
			={}& \frac{1}{(2\uppi)^2} \int_0^{\regulatorSC}\dr[p] p^2 \int_0^{\uppi} \dr[\theta] \sin \theta \frac{p q \cos \theta+q^2 + 2 \kron{i}{0} \mstar^2 }{2p q \cos \theta + q^2} \times \nonumber\\
			&\times\Bigg[ \frac{\OneMinusnsDef{E} }{ 2E }  -  \frac{\OneMinusnsDef{\E{\vecb{p+q}}}}{ 2\E{\vecb{p+q}}} \Bigg]\nonumber=\\
			\begin{split}
				={}&\frac{1}{8\uppi^2} \int_0^{\regulatorSC}\dr[p] p  
				\Bigg\{ 	\frac{\Theta( E^2 -\mu^2) }{ E} \left[\frac{q^2+\kron{i}{0}4 \mstar^2}{4q} \ln\left(\left|\frac{q+2p}{q-2p}\right|\right)+ p\right] + \\
				&  +\Theta\left( \E{ {p+q} }^2 -\mu^2\right) \Bigg[\frac{-1}{2 q}
				\begin{rcases}
					\begin{cases}
						\left(\E{p+q}-\mu\right)   &\ifc \mu^2\geq\E{p-q}^2 ,\\
						\left(\E{p+q}-\E{p-q}\right)  & \otherwisec. 
					\end{cases}
				\end{rcases}+  \\ 
				&  + \frac{q^2+\kron{i}{0} 4\mstar^2}{ 4 qE}
				\begin{cases}
					\ln\left|\frac{\mu^2-E^2}{2pq+q^2}\, \frac{(\E{p+q}+E)^2}{(|\mu|+E)^2} \right| &\ifc \mu^2\geq\E{p-q}^2 ,\\
					\ln \left|\frac{2p-q}{2p+q}\, \frac{(\E{p+q}+E)^2}{(\E{p-q}+E)^2}\right|   & \otherwisec .
				\end{cases}	
				\Bigg]\Bigg\}.
			\end{split}
		\end{align}
		At zero temperature, this yields
		\begin{align}
			&\EllTwo[,\phi_i]\left(q,\mstar,\mu,T=0\right) = \nonumber\\
			\begin{split}
				={}&\tfrac{1}{8\uppi^2}  \Bigg[\left(q^2+\kron{i}{0} 4\mstar^2\right) \left(\frac{E }{4 q}\ln \left| \tfrac{2 p+q}{2 p-q}\right| +\tfrac{1}{8} \ln \left| \tfrac{p+E}{p-E}\right| -\tfrac{\sqrt{4 \mstar^2+q^2}}{8 q} \ln \left| \tfrac{qE+p \sqrt{4 \mstar^2+q^2}}{qE-p\sqrt{4 \mstar^2+q^2} }\right| \right) +
				\\
				&\hphantom{\tfrac{1}{8\uppi^2}  \Bigg[}-\tfrac{1}{2} \left(p E-\mstar^2 \ln| \mstar p+\mstar E| \right)\Bigg]_{p=P_{\mathrm{L,0}}}^{p=\regulatorSC}+\\
				&+\tfrac{1}{8\uppi^2} 
				\Bigg[\frac{1}{2q}
				\begin{rcases}
					\begin{cases}
						-\frac{q\mstar^2\arsinh\left(\frac{p+q}{\left|\mstar\right|}\right)+q \left(p+q\right)\E{p+q}}{2}-\frac{\mu p^2}{2}+\frac{\E{p+q}^3}{3}  & \ifc \mu^2\geq\E{p-q}^2 ,\\
						-\frac{q\mstar^2 }{2}\left[\arsinh\left(\tfrac{p+q}{\left|\mstar\right|}\right)+\arsinh\left(\tfrac{p-q}{\left|\mstar\right|}\right)\right]+&\vphantom{\Bigg|} \\
						+\frac{2p^2-q^2+2\mstar^2}{6} (\E{p+q} -\E{p-q}) + \frac{qp}{6} (\E{p+q} +\E{p-q}) &\otherwisec  
					\end{cases}
				\end{rcases}+  \\ 
				&  + \frac{q^2+\kron{i}{0}4 \mstar^2}{4} \times 
				\\
				&
				\times \begin{rcases}
					\begin{cases}
						0 &\ifc \mu^2\geq\E{p-q}^2, \\
						\artanh \left(\frac{p}{E}\right)+\frac{E}{q} \ln \left|\frac{2p-q}{2p+q}\right| + \frac{\sqrt{q^2+4\mstar^2}}{2q} \ln \left|\frac{q E+\sqrt{q^2+4\mstar^2}p}{q E-\sqrt{q^2+4\mstar^2}p}\right|   &\otherwisec 
					\end{cases}	
				\end{rcases}\Bigg]_{p=P_{\mathrm{L,q}}}^{p=\regulatorSC}
				+
				\\
				&+\tfrac{1}{8\uppi^2}  \int_{P_{\mathrm{L,q}}}^{\regulatorSC} \dr p \,p \ \frac{q^2+\kron{i}{0} 4\mstar^2}{E 4 q}
				\begin{rcases}
					\begin{cases}
						\ln\left|\frac{\mu^2-E^2}{2pq+q^2}\,\frac{(\E{p+q}+E)^2}{(|\mu|+E)^2} \right| & \ifc \mu^2\geq\E{p-q}^2 ,\\
						\ln\left(\frac{(\E{p+q}+E)^2}{(\E{p-q}+E)^2}\right)\, ,  & \otherwisec
					\end{cases}	
				\end{rcases} 
			\end{split}, \label{eq:Appendix:NJLSCTZero}
		\end{align}
		and for $\mstar = 0$
		\begin{align}
			&\EllTwo[,\phi_i]\left(q,\mstar=0,\mu,T=0\right) = \\
			={}&\frac{1}{8\uppi^2}  \Bigg[\frac{q}{8} f(2p,q) +4 p^2\Bigg]_{p=\abs{\mu}}^{p=\regulatorSC}+\nonumber\\
			-&\frac{1}{16\uppi^2q} \left[\begin{rcases}
				\begin{cases}
					\frac{2}{3}p^3													&\ifc p\leq q-\abs{\mu} \\
					\frac{p^3}{3} + \frac{p^2}{2} (\abs{\mu}+q) -\frac{(\abs{\mu}-q)^2(5\abs{\mu}+q)}{6} &\ifc  q-\abs{\mu} < p < q+\abs{\mu}  \\
					p^2q + 3\mu^2 q - \frac{q^3}{3}&\ifc  p\geq q+\abs{\mu} 
				\end{cases}
			\end{rcases}\right]_{p=\max(0,\mu-q)}^{p=\regulatorSC} + \nonumber\\
			+& \tfrac{q}{32\uppi^2} \left[
			\begin{cases}
				\tfrac{1}{2} \left[f(2p,q)-4p\left(\ln q +1\right)\right]\vphantom{\Bigg|}&\ifc p\leq q-\abs{\mu}, \\
				\frac{q \ln |2p+q|}{2} -f(p,\abs{\mu}) + p \left(\ln \left(\frac{2p+q}{q }\right)-1\right)+ \vphantom{\Bigg|}&\\
				+ \tfrac{1}{2} \left[f(2(q-\abs{\mu}),q)-4(q-\abs{\mu})\left(\ln q +1\right)\right] +&\vphantom{\Bigg|}\\
				- \frac{q \ln |3q-\abs{\mu}|}{2} +f(q-\abs{\mu},\abs{\mu}) - (q-\abs{\mu}) \left(\ln \left(\frac{3q-\abs{\mu}}{q }\right)-1\right)  \vphantom{\Bigg|}&\ifc  q-\abs{\mu} < p < q+\abs{\mu},  \\
				\tfrac{f(2p,q)}{2} - \tfrac{f(2(q+\abs{\mu}),q)}{2} \\
				+ \tfrac{1}{2} \left[f(2(q-\abs{\mu}),q)-4(q-\abs{\mu})\left(\ln q +1\right)\right] +&\vphantom{\Bigg|}\\
				+ \frac{q \ln |3q+\abs{\mu}|}{2} -f(q+\abs{\mu},\mu) + (q+\abs{\mu}) \left(\ln \left(\frac{3q+\abs{\mu}}{q }\right)-1\right)+ \vphantom{\Bigg|}&\\
				- \frac{q \ln |3q-\abs{\mu}|}{2} +f(q-\abs{\mu},\abs{\mu}) - (q-\abs{\mu}) \left(\ln \left(\frac{3q-\abs{\mu}}{q }\right)-1\right)  \vphantom{\Bigg|}&\ifc  p\geq q+\abs{\mu} 
			\end{cases}
			\right]_{p=\max(0,\mu-q)}^{p=\regulatorSC}, \nonumber
		\end{align}
		where we defined
		\begin{align}
			f(x,y)= (x+y) \ln \abs{x+y}-(x-y) \ln \abs{x-y}.
		\end{align}
	\subsection{Lattice Regularization}
	\label{sec:Appendix:Latticestab}
	
		For the lattice regulations, one cannot split up $\Gamma^{(2)}$ according to \cref{eq:app:gamma2splitup}. 
		Instead, one needs to evaluate 
		\begin{align}
			\ell_{3, \phi_i}\left(\x{p},q,\mstar,\mu\right) ={}& \int_{-\regulatorLFT}^{\regulatorLFT}\!\!\!\! \intMeasureOverPiNew{p_0} \tr \left[\vertexV_i \quarkPropagatorSymbol_{X}\left(\X{p+(0,\x{q})}\right)\vertexV_i \quarkPropagatorSymbol_{X}\left(\X{p}\right)\right]  = \label{eq:Appendix:LatticeTracePropTemporalintegration} \\
			={}& \frac{\Nbar}{\sum_{i=1}^{3}\Big(\fermionDisp_X^2\left(\x{p_i}\right)-\fermionDisp_X^2\left(\x{p_i+q_i}\right)\Big)} \times \nonumber\\
			&\times\Bigg[\frac{E^2-\sum_{i=1}^{3}\fermionDisp_X\left(\x{p_i+q_i}\right) \fermionDisp_X\left(\x{p_i}\right) + \mstar^2}{E}  \left[\arctan\left(\tfrac{\regulatorLFT}{E-\abs{\mu}}\right)+\arctan\left(\tfrac{\regulatorLFT}{E+\abs{\mu}}\right)\right]+ \nonumber \\
			& \hphantom{\times\Bigg[}- \frac{E_{\x{p+q}}^2-\sum_{i=1}^{3}\fermionDisp_X\left(\x{p_i+q_i}\right) \fermionDisp_X\left(\x{p_i}\right) + \mstar^2}{E_{\x{p+q}}}   \left[\arctan\left(\tfrac{\regulatorLFT}{E_{\x{p+q}}-\abs{\mu}}\right)+\arctan\left(\tfrac{\regulatorLFT}{E_{\x{p+q}}+\abs{\mu}}\right)\right]\Bigg],
			\nonumber
		\end{align}
		with the lattice dispersion relations $\fermionDispSLAC$ and $\fermionDisp_{\text{Hybrid}}$ given in \cref{eq:5:SLACDisp} and \cref{eq:HybridDisp}, respectively, and $\vec{c}=( \mathds{1}, \iu \gamma_5 \tau_3)$. 
		The energy $E$ is defined as in \cref{eq:latt_energy} and 
		\begin{equation}
			E_{\x{p+q}} = \sqrt{\sum_{i=1}^3 \fermionDisp_X^2(p_i + q_i) + \mstar^2 } \label{eq:latt_energy_pplq}.
		\end{equation}
		The contribution of $\ell_{3, \phi_i}$ to $\gtwo[\phi_i]$ depends on the chosen lattice discretization and is given below.
	\subsubsection{SLAC Fermions}
	\label{sec:Appendix:SLACstab}
		For the SLAC discretization, the bosonic two-point function is then given by
		\begin{align}
			\gtwoArgs[\phi_i]{q}{\mstar}{\mu}{T=0} ={}& \frac{1}{2 \Gcoupling} - \frac{1}{2 \uppi}  \int_{-\regulatorLFT}^{\regulatorLFT} \!\!\!\! \intMeasureOverPiNew{p_1}\!\!\!\!
			\int_{-\regulatorLFT}^{\regulatorLFT} \!\!\!\! \intMeasureOverPiNew{p_2}\!\!\!\!
			\int_{-\regulatorLFT}^{\regulatorLFT} \!\!\!\! \intMeasureOverPiNew{p_3} \ell_{3, \phi_i}\left(\x{p},q,\mstar^2,\mu\right). \label{eq:Appendix:SLACTracePropTemporalintegration}
		\end{align}
	
	\subsubsection{Hybrid Fermions}
	\label{sec:Appendix:Hybridstab}
		
		For the Hybrid discretization, the bosonic two-point function is then given by
		\begin{align}
			\gtwoArgs[\phi_i]{q}{\mstar}{\mu}{T=0} ={}& \frac{1}{2 \Gcoupling} -  \weightFunctionFourier[X]\left(\x{q}\right) \weightFunctionFourier[X]\left(-\x{q}\right)  \frac{1}{2 \uppi} \int_{-\regulatorLFT}^{\regulatorLFT} \!\!\!\! \intMeasureOverPiNew{p_1}\!\!\!\!
			\int_{-\regulatorLFT}^{\regulatorLFT} \!\!\!\! \intMeasureOverPiNew{p_2}\!\!\!\!
			\int_{-\regulatorLFT}^{\regulatorLFT} \!\!\!\! \intMeasureOverPiNew{p_3} \ell_{3, \phi_i}\left(\x{p},q,\mstar^2,\mu\right)
			\label{eq:Appendix:HybridTracePropTemporalintegration}
		\end{align}
		with the weighting function $\weightFunctionFourier$ as introduced in Section \ref{sec:hybridFermions}.
		
	\section{Formulas for the Wave-Function Renormalization\label{app:formula_z}}
		According to \cref{eq:zsmall}, we perform a double derivative of $\gtwo[\phi_i]$ with respect to $q$ and take the limit $q\rightarrow 0$ in order to obtain $\z[\phi_i]$.
	
	\subsection{Pauli-Villars}
		For the \gls{pv} scheme, this amounts to 
		\begin{align}
			&\zArgs[\phi_i]{\mstar}{\mu}{T} =\nonumber\\
			\begin{split}
				={}& \frac{\Nbar}{16 \uppi^2} \pvsumck[0] \Bigg\{-\frac{1}{3} \frac{{\kron{i}{0}} \mstar^2}{\Mass_k^2} - \frac{1}{2} \ln \left(\frac{\Mass_k^2}{\regulatorPV^2}\right) - \int_0^{\infty} \dr[p] \frac{ \distributionF{E_k}+\distributionFBar{E_k}}{ E_k} +\\
				&\hphantom{\frac{\Nbar}{16 \uppi^2} \pvsumck[0] \Bigg\{}
				- \frac{\mstar^2}{3}\int_0^{\infty} \dr[p] \frac{1}{E_k^3}\left[  - \distributionF{E_k}-\distributionFBar{E_k} 
				+ \tfrac{E_k}{T}\left[\distributionF[2]{E_k} + \distributionFBar[2]{E_k} - \distributionF{E_k}-\distributionFBar{E_k}\right] \right] \Bigg\} \label{eq:app:zPVGeneral}
			\end{split}
		\end{align}
		and at zero temperature, this gives
		\begin{align}
			&\zArgs[\phi_i]{\mstar}{\mu}{T=0} = \nonumber\\
			={}& \frac{\Nbar}{16 \uppi^2} \pvsumck[0] \Bigg\{-\frac{1}{3} \frac{{\kron{i}{0}} \mstar^2}{\Mass_k^2} - \frac{1}{2} \ln \left(\frac{\Mass_k^2}{\regulatorPV^2}\right)+\nonumber\\
			&\hphantom{\frac{\Nbar}{16 \uppi^2} \pvsumck[0] \Bigg\{}
			-\int_0^{\infty} \dr[p] \frac{\Theta\left(\mu^2-E_k^2\right)}{E_k} - \frac{\mstar^2}{3}\int_0^{\infty} \dr[p] \frac{1} {E_k^3}\Big[  \Theta\left(\mu^2-E_k^2\right) + \frac{E_k}{\abs{\mu}} \delta \left(\tfrac{E_k}{\mu}-1\right) \Big] \Bigg\} \nonumber =\\
			={}&\frac{\Nbar}{16 \uppi^2} \pvsumck[0] \Bigg\{-\frac{1}{3} \frac{{\kron{i}{0}} \mstar^2}{\Mass_k^2} - \frac{1}{2} \ln \left(\frac{\Mass_k^2}{\regulatorPV^2}\right)  - \frac{ \Theta\left(\mub_k^2\right)}{3} \left[3 \arsinh\left(\frac{\mub_k}{\Mass_k}\right) +\frac{\mstar^2}{\Mass_k^2} \Abs{\frac{{\mub_k}}{ {\mu}}}+  \frac{ \mstar^2 }{ \mu\mub_k }  \right]\Bigg\}  \label{eq:app:zPVTZero}.
		\end{align}
		Evaluating \cref{eq:app:zPVGeneral} for $\mstar = 0$ gives
		\begin{align}
			&\zArgs[\phi_i]{\mstar=0}{\mu}{T} =\nonumber\\
			\begin{split}
				={}& \frac{\Nbar}{16 \uppi^2}  \Bigg\{\int_0^{\infty} \dr[p] \frac{1 }{p}\Big[1 -\distributionF{p}-\distributionFBar{p} + \tfrac{p}{T}\left(\distributionF[2]{p} + \distributionFBar[2]{p} - \distributionF{p}-\distributionFBar{p}\right)\Big] +\\
				&\hphantom{\frac{\Nbar}{16 \uppi^2}  \Bigg\{}
				+\pvsumck[1]\int_0^{\infty} \dr[p] \frac{p^2 }{E_k^3}\Big[1 - \distributionF{E_k}-\distributionFBar{E_k}  +\tfrac{E_k}{T}\left(\distributionF[2]{E_k} + \distributionFBar[2]{E_k}- \distributionF{E_k}-\distributionFBar{E_k}\right)\Big] \Bigg\} 	\label{eq:app:zPVMZero}
			\end{split}
		\end{align}
		and for $\mstar = T = 0$
		\begin{align}
			&\zArgs[\phi_i]{\mstar=0}{\mu\neq0}{T=0} = \nonumber\\
			\begin{split}
				={}&\frac{\Nbar}{16 \uppi^2} \Bigg\{\tfrac{1}{2} \ln (4) -\frac{1}{3}-\frac{\kron{i}{0}}{3} - \frac{1}{2} \ln \left(\frac{\mu^2}{\regulatorPV^2}\right)  +\\
				&\hphantom{\frac{\Nbar}{16 \uppi^2} }
				+ \pvsumck[1] \left[-\frac{1}{3} \frac{{\kron{i}{0}} \mstar^2}{\Mass_k^2} - \frac{1}{2} \ln \left(\frac{\Mass_k^2}{\regulatorPV^2}\right)  - \frac{ \Theta\left(\mub_k^2\right)}{3} \left[3 \arsinh\left(\frac{\mub_k}{\Mass_k}\right) +\frac{\mstar^2}{\Mass_k^2} \Abs{\frac{{\mub_k}}{ {\mu}}}+  \frac{ \mstar^2 }{ \mu\mub_k }  \right]\right] \Bigg\}. \label{eq:app:zPVMTZero}
			\end{split}
		\end{align}
	\subsection{Spatial Momentum Cutoff}
		In the \gls{sc} scheme, one finds
		\begin{align} 
			&\zArgs[\phi_i]{\mstar}{\mu}{T} =\label{eq:app:z3DCutoffGeneral}\\
			={}& \frac{\Nbar}{16 \uppi^2} \Bigg\{
			2 \int_0^{\regulatorSC} \dr[p] p^2 \frac{1}{E^3}\Big[1  - \distributionF{E}-\distributionFBar{E} 
			+ \tfrac{E}{T}\left(\distributionF[2]{E} + \distributionFBar[2]{E} -  \distributionF{E}-\distributionFBar{E}\right) \Big]
			+\nonumber\\
			&\hphantom{\frac{\Nbar}{16 \uppi^2} \Bigg\{}
			-3 \int_0^{\regulatorSC} \dr[p] \frac{\delta_{i,0}\, \mstar^2 p^2 + \frac{p^4}{3}}{E^5} \Bigg[1  -  \distributionF{E}-\distributionFBar{E} +\tfrac{E}{T}\left[\distributionF[2]{E} + \distributionFBar[2]{E}-  \distributionF{E}-\distributionFBar{E}\right] +\nonumber\\
			&\hphantom{\frac{\Nbar}{16 \uppi^2} \Bigg\{ -3 \int_0^{\infty} \dr[p] \frac{\mstar^2 p^2 + \frac{p^4}{3}}{E^5} \Bigg[}
			- \tfrac{E^2}{3T^2}  \left[ 2  \distributionF[3]{E} + 2\distributionFBar[3]{E} -3 \distributionF[2]{E} - 3 \distributionFBar[2]{E}+   \distributionF{E}+\distributionFBar{E}\right] \Bigg]+\nonumber\\
			&\hphantom{\frac{\Nbar}{16 \uppi^2} \Bigg\{}
			+\frac{\delta_{i,0}}{3} \int_0^{\regulatorSC} \dr[p]  \tfrac{5 \mstar^2 p^4 }{ E^7}  \Bigg[1  -  \distributionF{E}-\distributionFBar{E} +\tfrac{E}{T}\left[\distributionF[2]{E} + \distributionFBar[2]{E}-  \distributionF{E}-\distributionFBar{E}\right] +\nonumber\\
			&\hphantom{+\frac{2}{3} \int_0^{\infty} \dr[p]  \tfrac{5 \mstar^2 p^4 }{ E^7}  \Big[}
			- \tfrac{2E^2}{15T^2}  \left[ 2 \distributionF[3]{E} +  2 \distributionFBar[3]{E} - 3 \distributionF[2]{E} - 3 \distributionFBar[2]{E}+   \distributionF{E}+ \distributionFBar{E}\right]  +\nonumber\\
			&\hphantom{+\frac{2}{3} \int_0^{\infty} \dr[p]  \tfrac{5 \mstar^2 p^4 }{E^7}  \Big[}
			+ \tfrac{E^3}{15T^3}  \left[6  \distributionF[4]{E} + 6  \distributionFBar[4]{E} - 12 \distributionF[3]{E} -12\distributionFBar[3]{E} + 7 \distributionF[2]{E} +7   \distributionFBar[2]{E} -    \distributionF{E} -    \distributionFBar{E}\right]\Bigg]
			\Bigg\}. \nonumber 
		\end{align}
		Note that the higher orders of distribution functions compared to the \gls{pv} results stems from the fact that partial integration was not performed because the boundary terms would be non-vanishing.
		For $\mstar = 0$ one finds
		\begin{align}
			&\zArgs[\phi_i]{\mstar=0}{\mu}{T} =\nonumber\\
			={}& \frac{\Nbar}{16 \uppi^2} \Bigg\{
			\int_0^{\regulatorSC} \dr[p] \frac{1}{p}\Big[1  - \distributionF{p}-\distributionFBar{p} +
			+ \tfrac{p}{T}\left[\distributionF[2]{p} + \distributionFBar[2]{p} -  \distributionF{p}-\distributionFBar{p}\right] \Big]
			+\nonumber\\
			&\hphantom{\frac{\Nbar}{16 \uppi^2} \Bigg\{}
			+ \int_0^{\regulatorSC} \dr[p] \tfrac{p}{3T^2} 
			\left[
			2  \distributionF[3]{p} + 2\distributionFBar[3]{p} -3 \distributionF[2]{p} - 3 \distributionFBar[2]{p}+   \distributionF{p}+\distributionFBar{p}\right]
			\Bigg\} .  \label{eq:app:z3DMZero}
		\end{align}

\end{document}